\documentclass[useAMS,usenatbib]{mnras}

\usepackage{tabto}
\usepackage{float}
\usepackage{color}
\usepackage{amssymb} % Permite utilizar outros simbolos matematicos
\usepackage{amsfonts,amsmath,amscd} % Permite usar simbolos matematicos
\usepackage[subnum]{cases}
\usepackage{mathtools}
\usepackage{verbatim}
\usepackage{t1enc}
\usepackage[T1]{fontenc}
\usepackage{ae,aecompl}
\usepackage[dvips]{graphics}
\usepackage{adjustbox}
\usepackage{eufrak} % Permite utilizar estilos diferentes de letra
\usepackage{euscript} % Permite utilizar estilos diferentes de letra
\usepackage{graphicx} % Permite inserir figuras
\usepackage[normalem]{ulem} % Permite mais formatacao do texto
\usepackage{makeidx} %Produz indice remissvo (glossario)
\usepackage[usenames,dvipsnames]{pstricks}
\usepackage{epsfig}
\usepackage{pst-grad} % For gradients
\usepackage{pst-plot} % For axes
\usepackage{pstcol,times,graphics,pstricks,pst-node,pst-coil} % Permite desenhar
\usepackage{multicol} % Allows for multiple columns
\usepackage{multirow} % Permite linhas multiplas
\usepackage{indentfirst} % Permite indentar o Primeiro paragrafo de uma secao
\usepackage{natbib} % Estilo de HARVARD
\usepackage{setspace} % Define espacamento
\usepackage{threeparttable}
\newcommand{\specialcell}[2][c]{%
  \begin{tabular}[#1]{@{}c@{}}#2\end{tabular}}

\title[CVs in MOCCA GCs: CAML and CV evolution]
{MOCCA-SURVEY database I. Accreting white dwarf binary systems in globular clusters -- III. Cataclysmic variables -- Implications of model assumptions}
\author[Belloni et al.]
{Diogo Belloni$^{1,2}$\thanks{E-mail: belloni@camk.edu.pl (DTB)},
M{\'o}nica Zorotovic$^3$, Matthias R. Schreiber$^{3,4}$,
\newauthor Nathan W. C. Leigh$^{5}$, Mirek Giersz$^1$ and Abbas Askar $^{1}$\\
$^{1}$ Nicolaus Copernicus Astronomical Centre, Polish Academy of Sciences, ul. Bartycka 18, PL-00-716 Warsaw, Poland \\
$^{2}$ CAPES Foundation, Ministry of Education of Brazil, DF 70040-020, Brasilia, Brazil \\
$^{3}$ Instituto de F{\'i}sica y Astronom{\'i}a, Universidad de Valpara{\'i}so, Av. Gran Breta{\~n}a 1111 Valpara{\'i}so, Chile \\
$^{4}$ Millenium Nucleus ``Protoplanetary Disks in ALMA Early Science'', Universidad de Valpara{\'i}so, Valpara{\'i}so 2360102, Chile \\
$^{5}$ Department of Astrophysics, American Museum of Natural History, Central Park West and 79th Street, New York, NY 10024, USA}
\begin{document}
%%%%%%%%%%%%%%%%%%%%%%%%%%%%%%%%%%%%%%%%%%%%%%%%

\date{Accepted 2017 March 2. Received 2017 March 1; in original form 2017 February 6}

\pagerange{\pageref{firstpage}--\pageref{lastpage}} \pubyear{2016}

\maketitle

\label{firstpage}

%%%%%%%%%%%%%%%%%%  ABSTRACT %%%%%%%%%%%%%%%%%%
\begin{abstract}
In this third of a series of papers related to cataclysmic variables (CVs) and
related objects, we analyse the population of CVs in a set of 12 globular cluster models evolved 
with the MOCCA Monte Carlo code, for two initial binary populations (IBPs),
two choices of common-envelope phase (CEP) parameters, and three different models for the 
evolution of CVs and the treatment of angular momentum loss. 
When more realistic models and parameters are considered, we find that 
present-day cluster CV duty cycles are extremely-low ($\lesssim 0.1$ per cent) which
makes their detection during outbursts rather difficult. Additionally, the IBP
plays a significant role in shaping the CV population properties, and models that
follow the Kroupa IBP are less affected by enhanced angular momentum loss. We also predict
from our simulations that CVs formed dynamically in the past few Gyr (massive CVs) 
correspond to bright CVs (as expected), and that faint CVs formed several Gyr ago 
(dynamically or not) represent the overwhelming majority.
Regarding the CV formation rate, we rule out the notion that it is similar
irrespective of the cluster properties.
Finally, we discuss the differences in the present-day CV properties related to the 
IBPs, the initial cluster conditions, the CEP parameters, formation channels, 
the CV evolution models, and the angular momentum loss treatments.
\end{abstract}


%%%%%%%%%%%%%%%%%%%%%%%%%%%%%%%%%%%%%%%%%%%%%%%%%%%%%%%%%%%%%%%%%%
% NEW SECTION
%%%%%%%%%%%%%%%%%%%%%%%%%%%%%%%%%%%%%%%%%%%%%%%%%%%%%%%%%%%%%%%%%%
\section{INTRODUCTION}

Cataclysmic variables (CVs) are interacting
binaries composed of a white dwarf (WD) undergoing stable 
mass transfer from a main sequence (MS) star or a brown dwarf (BD) 
\citep[e.g.][]{Warner_1995_OK,Knigge_2011_OK}.  They are 
expected to exist in non-negligible numbers in globular clusters
(GCs) that are natural laboratories for testing theories of stellar 
dynamics and evolution. 

CVs in GCs have been studied by many authors, both 
theoretically and observationally 
\citep[e.g.][and references therein]{Knigge_2012MMSAI}. GCs 
are thought to play a crucial role in CV formation, since their 
densities are sufficiently high that dynamical encounters 
involving binaries should be common. Thus, in dense 
GCs, it is natural to expect that many CV progenitors will have 
been affected by dynamics in some way prior to CV formation 
\citep[e.g.][]{Shara_2006,Ivanova_2006,Belloni_2016a,Belloni_2016b,Hong_2016}. 

In the first paper of this series \citep{Belloni_2016a}, we discussed 
six specific MOCCA (MOnte Carlo Cluster simulAtor) models with a focus on the properties of
their present-day CV populations. In the second paper \citep{Belloni_2016b}, 
we concentrated instead on a discussion of the properties of the progenitor 
and formation-age populations, and how CV properties are affected
by their ages. 

Here we aim to complement our previous work by considering the same six 
initial cluster conditions that we already presented but considering one new combination
of the common-envelope phase (CEP) parameters and two new approaches 
to the treatment of CV evolution and angular momentum loss (AML). We analyze the impact 
of these variables on present-day CV populations.

The paper is structured as follows. In Section \ref{model}, we describe 
the codes used for the cluster and CV evolution simulations.
We present the suite of models analyzed in this paper in Section \ref{models}. 
The main results of this investigation are presented in Sections \ref{gc_evol_cep_par},
\ref{fap_formation_rate}, \ref{number_pdp_cvs}, and \ref{pdp},
while Section \ref{discussion} contains a discussion of the main results, along with a direct comparison 
between our simulated CV properties and observed CV candidates in GCs. Finally, we summarize our 
main conclusions in Section \ref{conclusions}.

%%%%%%%%%%%%%%%%%%%%%%%%%%%%%%%%%%%%%%%%%%%%%%%%%%%%%%%%%%%%%%%%%%
% NEW SECTION
%%%%%%%%%%%%%%%%%%%%%%%%%%%%%%%%%%%%%%%%%%%%%%%%%%%%%%%%%%%%%%%%%%

\section{METHODOLOGY}
\label{model}

We start with a summary of the main aspects of the codes used in 
\citet[][]{Belloni_2016a,Belloni_2016b}, defining as well the different dynamical
groups and the different populations that we will analyze. 
In Section \ref{cv_evolution} we explain how and why we introduced a new code
for simulating close WD-MS binaries and CV evolution, which is up-to-date and current. 
Finally, in Section \ref{CAML} we describe the three approaches adopted in this 
work to treat the AML during the CV phase.

The cluster evolution is performed with the MOCCA code, which is based 
on the orbit-averaged Monte Carlo technique from \citet{Henon_1971} 
improved by \citet{Stodolkiewicz_1986} and further improved by 
\citet[][and references therein]{Giersz_2013}. These 
last authors included in MOCCA the FEWBODY code \citep{Fregeau_2004} 
to handle 3 and 4-body interactions, a description of escape 
processes in tidally limited clusters \citep{Fukushige_2000}, and stellar evolution via
the SSE/BSE code developed by \citet{Hurley_2000,Hurley_2002}. Additionally, 
MOCCA has been extensively tested against N-body codes \citep[e.g.][]{Giersz_2013,Wang_2016}, 
showing good agreement.
%DB: I put more info because some of them are important. For example, the reviewer of our
%first paper asked about info regarding comparison between MOCCA and N-body codes. This
%might happen again.

The analysis of the CVs produced in clusters simulated by the MOCCA code was performed with
the CATUABA ({\bf C}ode for {\bf A}nalysing and s{\bf TU}dying c{\bf A}taclysmic 
variables, sym{\bf B}iotic stars and {\bf A}M CVns) code.

A detailed description of the MOCCA and CATUABA codes, as well as all 
the features and main assumptions adopted in them, can be found in \citet[][]{Belloni_2016a,Belloni_2016b}. 

%%%%%%%%%%%%%%%%%%%%%%%%%%%%%%%%%%%%%%%%%%%%%%%%%%%%%%%%%%%%%%%%%%
% NEW SECTION
%%%%%%%%%%%%%%%%%%%%%%%%%%%%%%%%%%%%%%%%%%%%%%%%%%%%%%%%%%%%%%%%%%

\subsection{The influence of dynamics on CV formation}
\label{catuaba_dyn}
%DB: I merged the two subsections.

As already explained in \citet[][see section 3.2, for more details]{Belloni_2016a}, 
%DB: I replaced the reference and removed that just below the "\end{itemize}"
CATUABA permits for the unambiguous identification of a given CV population
in the cluster snapshots provided by MOCCA at various times in the cluster evolution.  
Based on the most relevant events in the history of each individual CV, systems are
separated into three groups: 
\begin{itemize}
\item {\it Binary stellar evolution} (BSE) group, which corresponds to CVs formed without 
any influence from dynamics. 
\item {\it Weak dynamical interactions} (WDI) group, i.e. CVs weakly influenced by dynamics. 
\item {\it Strong dynamical interaction} (SDI) group, which contains those CVs strongly 
influenced by dynamical interactions, including systems that underwent either exchanges 
or mergers.
\end{itemize}

The CATUABA code also recognizes the main formation channels for the WD-MS 
population that will later on become the CV population. Briefly, close WD-MS binaries 
can be formed through a CEP with/without weak/strong dynamical interactions, through 
exchanges with/without a CEP, and through mergers with/without a CEP.

%%%%%%%%%%%%%%%%%%%%%%%%%%%%%%%%%%%%%%%%%%%%%%%%%%%%%%%%%%%%%%%%%%
% NEW SECTION
%%%%%%%%%%%%%%%%%%%%%%%%%%%%%%%%%%%%%%%%%%%%%%%%%%%%%%%%%%%%%%%%%%

%\subsection{CV populations}
%\label{catuaba_pop}
%DB: I removed this subsection. It seems that its content works better
%in the above one.

Three different  populations can also be distinguished:
\begin{itemize}
\item The \textit{progenitor population} is defined as the population of all binaries that are CV progenitors, 
i.e. the population of binary systems at the time of cluster birth (i.e., $t=0$) that will become CVs.  
%DB: I added "binary" and removed the apostrophes in the word CV.
\item The \textit{formation-age population} is defined as the population of zero-age CVs, i.e. when mass 
transfer starts from the donor onto the WD.  Obviously, the time corresponding to the onset of 
the CV phase is different for every CV.  
\item The \textit{present-day population} is the population of CVs at the current age (t $=$ 12 Gyrs).
\end{itemize}

%%%%%%%%%%%%%%%%%%%%%%%%%%%%%%%%%%%%%%%%%%%%%%%%%%%%%%%%%%%%%%%%%%
% NEW SECTION
%%%%%%%%%%%%%%%%%%%%%%%%%%%%%%%%%%%%%%%%%%%%%%%%%%%%%%%%%%%%%%%%%%

\subsection{Close WD-MS binary evolution and CV evolution}
\label{cv_evolution}

The most important part of MOCCA for producing CVs and related
objects is the Binary Star Evolution (BSE) code from \citet{Hurley_2002},
which models angular momentum loss in binaries.  However, its 
implementation is not fully up-to-date 
\citep[][for a state-of-the-art model]{Knigge_2011_OK,Schreiber_2016}. 
Mass transfer occurs if either star fills its Roche lobe and 
may proceed on a nuclear, thermal, or dynamical time-scale. 
Prescriptions to determine the type and rate of mass
transfer and the response of the primary to accretion 
are implemented in BSE. 
For the CEP, BSE assumes that the common envelope
binding energy is that of the giant(s) envelope involved
in the process. 

The overall CV evolution can be recovered by BSE, although the
CV evolution model implemented in this code is out-dated 
(for more details, see Section \ref{discussion_CVs_aml}).
This leads to unrealistic predictions for the location
of the period minimum and the subsequent CV evolution. Additionally,
BSE is unable to reproduce the orbital period gap, which is one of the
most important observables related to CVs.

Since the BSE code is out-dated with respect to many aspects
of CV evolution, we use instead a different
code to simulate the entire population of close WD-MS
binaries formed in the simulated clusters. 
For this purpose, we implement in CATUABA an
option to extract all close WD-MS binaries (with periods less
than 10 days) that are formed throughout the cluster evolution.
The systems are then evolved up to 12 Gyr with the
code described in \citet{Zorotovic_2016}.

We emphasize that in order to proceed with this method, 
it is reasonable to assume that once a close WD-MS binary 
has formed then its subsequent evolution will not be influenced 
by dynamics based on the results of \citet{Belloni_2016b}, who showed that CVs
are very dynamically hard, and of low mass, such that the 
%NL: Don't you mean that the impact parameter is large?  I changed this below, so if I am wrong please do change it back.
%DB: I think the impact parameter is smaller, which causes less interactions. No? I changed back.
impact parameter for interactions is small. 
It follows that the probability of an interaction occurring 
is very small \citep[see ][for more details regarding the interruption of 
binary mass transfer due to dynamical interactions]{Leigh_2016}. 
Thus, assuming that the evolution of close 
WD-MS binaries is not affected by dynamics is not far 
from what happens in simulations of realistic clusters.

%%%%%%%%%%%%%%%%%%%%%%%%%%%%%%%%%%%%%%%%%%%%%%%%%%%%%%%%%%%%%%%%%%
% NEW SECTION
%%%%%%%%%%%%%%%%%%%%%%%%%%%%%%%%%%%%%%%%%%%%%%%%%%%%%%%%%%%%%%%%%%

\subsection{Angular Momentum Loss and stability limit}
\label{CAML}

It is well-known that AML in a CV determines its secular evolution. 
It is important to distinguish 
AML due to mass transfer, usually called consequential angular
momentum loss (CAML, $\dot{J}_{\rm CAML}$) from AML independent of 
mass transfer (e.g. magnetic braking, gravitational radiation), usually 
called systemic AML ($\dot{J}_{\rm sys}$) \citep[e.g.][]{King_1995}.
Typical candidates for the CAML are: circumbinary disks \citep[e.g.][]{Willems_2005}, 
hydromagnetic accretion disk winds \citep[e.g.][]{Cannizzo_1988}, 
and frictional drag associated with nova eruptions \citep{Schreiber_2016,Nelemans_2016}.

Particularly interesting is the influence of nova eruptions on CV
evolution. According to \citet{Schreiber_2016}, if the strength of 
CAML is inversely proportional to the WD mass, then most CVs with WD masses
less than $\sim$ 0.5 ${\rm M_\odot}$ are dynamically unstable.
The motivation for such a functional form (empirical CAML) comes from the fact
that a helium (He) WD has never been observed in a CV in the Galactic field, although
such WDs are observed in detached systems \citep{Zorotovic_2011}.
The primary mechanism thought to be responsible for such a dependence between CAML and WD mass
is nova eruptions \citep{Schreiber_2016,Nelemans_2016}. 
The frictional AML produced by novae depends strongly 
on the expansion velocity of the ejecta \citep{Schenker_1998}. For low-mass WDs, the
expansion velocity is small \citep{Yaron_2005}, and this may lead 
to strong AML by friction that makes most CVs with He WDs dynamically unstable.

%At this point we must say that BSE assumes that only 0.1 per cent of the
%transferred matter is, in fact, accreted by the WD \citep[][Eq. 66]{Hurley_2002}. 
%Due to nova eruptions, most of the accreted material is expelled. On the other hand, 
%the code described in \citet{Zorotovic_2016} assumes that
%all accreted matter is expelled during nova eruptions, which is coherent
%as shown by \citet{Schenker_1998} for the secular evolution of CVs.
%In addition, in both codes, there is no separation of CVs with
%respect to classical novae (CNe) and dwarf novae (DNe), where CNe have
%much higher mass transfer rates. However, a transition from CN
%to DN (and vice-versa) is expected since recently a few old
%nova shells have been observed containing a DN inside \citep[e.g.][]{Shara_2007} 
%which clearly indicates that CNe become DNe. Finally, \citet{Schmidtobreick_2015}
%analysed 15 DNe and showed that the recurrence time of the CN cycle is at least 
%13000 years, which greatly lengthens the lifetimes of old novae.

%NL: I think a citation and/or some further explanation to the below is needed.  What "problems", specifically?
%DB: Monica added the reference to this point, i.e. Schreiber et al. (2016)

The empirical CAML model is a good candidate 
to solve several problems related to CV evolution, like the missing low-mass WDs in CVs, 
the predicted versus observed space density or the period distribution 
\citep[see][for more details]{Schreiber_2016}. Recently, this model has also
been proposed as the explanation for the existence of single He-core WDs \citep{Zorotovic_2017}.
Therefore, and despite the fact that the physical mechanism behind the
new empirical form of CAML is not understood at a fundamental level, we think 
it is appropriate to check the influence that this enhanced CAML can have on 
GC CVs.

Both codes used in this work for CV evolution do not follow the evolution of the mass transfer rate
during nova cycles but just calculate the secular mean mass transfer rate. Mass transfer variations 
that may occur on shorter time scales between two nova eruptions \citep[e.g.][]{Shara_1986,Shara_2007,Schmidtobreick_2015} 
are thus not considered. As shown by \citet{Schenker_1998} this should not affect the stability limit 
which is fundamental for the predictions of binary population models. In both codes the mass loss due 
to nova eruption is simply taken into account by assuming that only a small fraction of the transferred 
mass is accreted by the WD. In BSE this fraction is 0.1 per cent \citep[][Eq. 66]{Hurley_2002} so the 
WD mass is assumed to slowly grow. \citet{Zorotovic_2016} assume that all accreted matter is expelled 
during nova eruptions i.e. the WD mass is kept constant.

Both the BSE code and the code described in \citet{Zorotovic_2016} use systemic AML  
given by the disrupted magnetic braking model \citep{Rappaport_1983,Schreiber_2010} for the evolution 
of close detached binaries. This model assumes that magnetic braking is not active, 
or at least strongly inefficient, in systems with fully convective MS secondary stars.
However, when mass transfer starts, \citet{Zorotovic_2016} include CAML as an extra
source of AML, which affects the orbital period evolution and the critical mass ratio for 
stable mass transfer. In BSE, on the other hand, CAML is assumed to be negligible and
therefore is not included. 

In order to check the influence of CAML on CVs in GCs, in addition
to those CVs formed in MOCCA, we simulate all close WD-MS formed 
in our models with the code described in \citet{Zorotovic_2016}.
We consider in this investigation two formulations for CAML, namely
the classical CAML (cCAML) from \citep{King_1995} and the empirical 
formulation (eCAML) from \citet{Schreiber_2016}, given below:

\begin{equation}
\frac{\dot{J}_{\rm CAML}}{J} \ = \ \nu \frac{\dot{M}_{2}}{M_{2}}
\end{equation}
where

\begin{equation*}
    \nu \ = \
  \begin{dcases}
    \frac{M_{2}^2}{M_{1}(M_{1} +M_{2})},	& \text{classical \citep{King_1995}} \\ 
				  	    	& \\
    \frac{0.35}{M_{1}}, 	  		& \text{empirical \citep{Schreiber_2016}} 
  \end{dcases}
\end{equation*}

\

\

For both CAML approaches, we compute the stability limit imposed
by the adopted formulation by equating the adiabatic mass radius 
exponent and the mass radius exponent of the secondary's Roche lobe. 
In BSE, on the other hand, a fixed value for the critical mass ratio is 
assumed.

%%%%%%%%%%%%%%%%%%%%%%%%%%%%%%%%%%%%%%%%%%%%%%%%%%%%%%%%%%%%%%%%%%
% NEW SECTION
%%%%%%%%%%%%%%%%%%%%%%%%%%%%%%%%%%%%%%%%%%%%%%%%%%%%%%%%%%%%%%%%%%

%\subsection{Summary of Methodology}
%\label{sum_method}

%%%%%%%%%%%%%%%%%%%%%%%%%%%%%%%%%%%%%%%%%%%%%%%%%%%%%%%%%%%%%%%%%%
% NEW SECTION
%%%%%%%%%%%%%%%%%%%%%%%%%%%%%%%%%%%%%%%%%%%%%%%%%%%%%%%%%%%%%%%%%%

\section{MODELS}
\label{models}

In what follows we describe the main initial conditions assumed for the GCs
in our different models. The most important differences concern 
the initial binary population (IBP), initial binary fraction 
and initial central density.
We also explain our choice to introduce a new set of CEP parameters
in order to generate six new models. 

\subsection{Initial cluster conditions}
%DB: I replaced the title of this subsection.
The IBP refers to all initial binaries in a given cluster, 
associated with specific distributions for their orbital parameters. 
In this work, we continue the analysis started in 
\citet{Belloni_2016a,Belloni_2016b}, who investigated models with two distinct
IBPs:
\begin{itemize}
\item K models: models with 95 per cent primordial binary fractions 
with orbital parameter distributions that follow the Kroupa IBP,
which is based on binary eigenevolution and a mass feeding algorithm 
\citep{Kroupa_INITIAL}.
\item S models: models with low binary fractions of either 5 or 10 
per cent that follow the ``Standard'' initial binary orbital 
parameter distributions. The Standard IBP assumes an 
uniform distribution for the mass ratio, an uniform distribution 
in the logarithm of the semi-major axis $\log(a)$ or a log-normal 
semi major axis distribution, and a thermal eccentricity distribution. 
%DB: I added the "log-normal" options for semi-major axis.
\end{itemize}

The main distributions associated with both IBPs are shown in Figures 
\ref{A1} and \ref{A2} in the Appendix \ref{ap}.
%DB: I added the reference to the appendix.
%NL: Al of the new edits look great so far!

The models also differ with respect to the initial central density. We consider:
sparse ($\rho_c \sim 10^3$ M$_\odot$ pc$^{-3}$), 
dense ($\rho_c \sim 10^5$ M$_\odot$ pc$^{-3}$), 
and very dense ($\rho_c > 10^5$ M$_\odot$ pc$^{-3}$) clusters,
where $\rho_c$ is the cluster central density.  

%Additionally, we adopt two different initial mass functions (IMFs) that follow a 
%broken power law $\xi(m) \propto m^{-\alpha}$, defined in 
%\citet{Kroupa_1991,Kroupa_1993}. The range of stellar masses assumed 
%in this study is between $0.08 {\rm M_\odot}$ and $100 {\rm M_\odot}$. 
%MIREK: I removed the above.

We assume that all stars are on the zero-age main sequence
when the simulations begin, and that all residual gas left over from the star
formation process has already been expelled from the cluster. Additionally,
all models have low metallicity ($\lesssim$ 0.001), are initially in virial equilibrium, and have neither
rotation nor primordial mass segregation \citep[e.g.][]{Leigh_2013,Leigh_2015}. Finally, all models are evolved 
for 12 Gyr, up to the present-day.

The properties of the six models are summarized in Table \ref{Tabmodels}. 

\begin{table*} 
\centering
\caption{Models and the parameters that define them. The name of each model has a letter and a number.
The letter indicates its initial binary population as well as its initial binary fraction,
namely: K (Kroupa) with high initial binary fraction and S (Standard) with low initial binary fraction
(see the text); and the number indicates its central density: 1 (sparse), 2 (dense) and 3 (very dense).}
\label{Tabmodels}
\begin{adjustbox}{max width=\textwidth}
\noindent
\begin{threeparttable}
\noindent
\begin{tabular}{l|c|c|c|c|c|c|c|c|c|c|c|c|c|c}
\hline\hline
Model & \specialcell{Mass\\$\left[ {\rm M}_\odot \right]$} & \specialcell{Number of\\objects}   &  \specialcell{Initial\\binary fraction} & \specialcell{Central Density\\$\left[ {\rm M}_\odot \; {\rm pc}^{-3} \right]$} &  \specialcell{$r_{\rm t}$\\$[{\rm pc}]$}  & \specialcell{$r_{\rm h}$\\$[{\rm pc}]$} & Z & IMF \tnote{a} & q \tnote{b,c} & a \tnote{b,c} & e \tnote{b,c} & \specialcell{Present-day\\Type \tnote{d}}  & \specialcell{Present-day\\binary fraction} \\ 
\hline\hline
K1 & \hspace{0.2cm} $8.07 \times 10^5$ & \hspace{0.2cm} $1.12 \times 10^6$ &  \hspace{0.2cm} $95$ \% & \hspace{0.2cm} $1.9 \times 10^2$ & \hspace{0.2cm} $115$ & \hspace{0.2cm} $16.9$ & $0.001$ & $3$ & Kroupa & Kroupa & Kroupa & pc & $28.3$ \% \\ \hline
K2 & \hspace{0.2cm} $8.07 \times 10^5$ & \hspace{0.2cm} $1.12 \times 10^6$ &  \hspace{0.2cm} $95$ \% & \hspace{0.2cm} $7.8 \times 10^4$ & \hspace{0.2cm} $115$ & \hspace{0.2cm} $2.3$ & $0.001$ & $3$ & Kroupa & Kroupa & Kroupa & cIMBH & $8.9$ \% \\ \hline

K3 & \hspace{0.2cm} $8.07 \times 10^5$ & \hspace{0.2cm} $1.12 \times 10^6$ &  \hspace{0.2cm} $95$ \% & \hspace{0.2cm} $3.2 \times 10^5$ &  \hspace{0.2cm} $72$ & \hspace{0.2cm} $1.4$ & $0.001$ & $3$ & Kroupa & Kroupa & Kroupa & cIMBH & $5.4$ \% \\ \hline
S1 & \hspace{0.2cm} $5.92 \times 10^5$ & \hspace{0.2cm} $1.00 \times 10^6$ &  \hspace{0.2cm} $05$ \% & \hspace{0.2cm} $2.8 \times 10^3$ &  \hspace{0.2cm} $97$ &  \hspace{0.2cm} $7.5$ & $0.00016$ & $3$ & uniform & log-normal & thermal & c & $4.9$ \% \\ \hline
S2 & \hspace{0.2cm} $9.15 \times 10^5$ & \hspace{0.2cm} $1.80 \times 10^6$ &  \hspace{0.2cm} $10$ \% & \hspace{0.2cm} $1.3 \times 10^5$ & \hspace{0.2cm} $125$ &  \hspace{0.2cm} $2.1$ & $0.001$ & $2$ & uniform & \specialcell{uniform\\in $\log(a)$}  &  thermal & pc & $4.8$ \% \\ \hline
S3 & \hspace{0.2cm} $1.52 \times 10^5$ & \hspace{0.2cm} $3.00 \times 10^5$ &  \hspace{0.2cm} $10$ \% & \hspace{0.2cm} $5.9 \times 10^5$ &  \hspace{0.2cm} $69$ &  \hspace{0.2cm} $0.7$ & $0.001$ & $2$ & uniform & \specialcell{uniform\\in $\log(a)$} & thermal & c & $4.6$ \% \\ \hline \hline
\end {tabular}
\begin{tablenotes}
       \item[a] The IMF 3 is the Kroupa IMF with three segments \citep{Kroupa_1993} and the IMF 2 is the Kroupa IMF with two segments \citep{Kroupa_1991}.
       \item[b] The Kroupa IBP corresponds to the construction of the parameters based on the eigenevolution and the mass feeding algorithm \citep{Kroupa_INITIAL}.
       \item[c] The Standard IBP is associated with an uniform distribution for the mass ratio, a log-uniform or log-normal distribution for the semi-major axis and a thermal eccentricity distribution.%  These distributions follow from an uniform binding-energy distribution.
       \item[d] The cluster present-day type can be: post-core collapse (c), post-core collapse with intermediate-mass black hole (cIMBH) and pre-core collapse (pc).
\end{tablenotes}
\end{threeparttable}
\end{adjustbox}
\end{table*}

%%%%%%%%%%%%%%%%%%%%%%%%%%%%%%%%%%%%%%%%%%%%%%%%%%%%%%%%%%%%%%%%%%
% NEW SECTION
%%%%%%%%%%%%%%%%%%%%%%%%%%%%%%%%%%%%%%%%%%%%%%%%%%%%%%%%%%%%%%%%%%

\subsection{Common-envelope phase}
\label{int_CEP}

An important phase in the life of most interacting binaries
(such as CVs, AM CVns, X-ray binaries, etc.)
is the CEP during which time two stars share the same envelope  
\citep[see ][for a review on common-envelope evolution]{Ivanova_REVIEW}. 
This phase leads to binaries whose orbital periods are orders of 
magnitude shorter than prior to this phase, due to common envelope 
ejection and gas dynamical friction.
 
Although a great deal of effort has been put
in to understanding the CEP, the energy budget 
during this phase is not well understood, especially
the role of recombination energy stored in ionized
regions inside the envelope \citep{Ivanova_2015}.

For convenience, the outcome of this phase is usually approximated 
by the balance between the change in the orbital energy and
envelope binding energy, given an efficiency $\alpha$ 
for this process. More specifically, $\alpha$ is the efficiency 
with which orbital energy is used to eject the envelope, i.e.
it corresponds to the fraction of the difference in orbital 
energy (before and after the CE) in unbinding the envelope, i.e.

%NL: Why is the \Delta here?  Doesn't \alpha account for the fact that this is a fraction?
%MZ: No, it is a fracion of the difference (delta) in orbital energy before and after the CEP. You don't loose all the orbital energy, unless they merge during the CEP

\begin{equation}
E_b \ = \ \alpha \Delta E_{\rm orb} \label {EQ1}
\end{equation}
where $E_b$ is the envelope binding energy, $E_{\rm orb}$ is the
orbital energy, and $\alpha$ is the fraction of $E_{\rm orb}$
used to unbind the envelope (i.e. CEP efficiency).
The binding energy is usually approximated by
\begin{equation}
E_b \ = \ \frac{GM(M-M_c)}{\lambda R}. \label{EQ2}
\end{equation}
where $\lambda$ is the binding energy parameter, which depends on 
the structure of the primary star.

If a fraction $\alpha_{\rm rec}$ of the recombination energy
of the envelope is used to help in ejecting the envelope,
then the binding energy equation is

\begin{equation}
E_b \ = \ - \int_{M_c}^M \frac{G \; m}{r(m)} \; dm + \alpha_{\rm rec} \int_{M_c}^M \varepsilon_{\rm rec}(m)dm \label{EQ3}
\end{equation}
where $M_c$ is the giant core mass, $M$ is the giant mass, $\varepsilon_{\rm rec}$
is the specific recombination energy, and $\alpha_{\rm rec}$ is the fraction of 
the recombination energy used to unbind the envelope. 
The binding energy parameter $\lambda$ can be computed by equating 
equations \ref{EQ2} and \ref{EQ3}. When $\alpha_{\rm rec} > 0$, 
an additional source of energy is used in the ejection
of the envelope, which implies that less orbital energy 
is needed to unbind the envelope.

In the six models investigated by \citet{Belloni_2016a,Belloni_2016b},
we assumed $\alpha=3$, which might not be an appropriate
choice based on the study carried out by \citet{Zorotovic_2010}. These authors 
analyzed a homogeneous sample of 60 post common envelope binaries (PCEBs) identified
by the Sloan Digital Sky Survey. They found that
$\alpha$ should be $\sim$ 0.2-0.3, and certainly no greater than 1.
In our previous works, we also used $\lambda=0.5$. As pointed out by \citet{Zorotovic_2014},
in the original version of BSE, a fixed value for the binding energy 
parameter ($\lambda$) is assumed as input to the code. However, after 
improvements were made, a function to compute $\lambda$ was implemented 
in BSE, as described in detail by \citet[][see their Appendix A]{Claeys_2014}. 
In this newer implementation, a positive value of $\lambda$
represents the fraction of the recombination energy 
($\alpha_{\rm rec}$) included in the calculation of the
the binding energy parameter. On the other hand, for 
$\lambda = 0$, the binding energy parameter is computed
as a function of the WD progenitor (without any additional
source of energy). For $\lambda < 0$, the 
binding energy parameter is fixed and set equal to -$\lambda$.
Therefore, using $\lambda=0.5$ as an input for the code, we were
in fact assuming that 50\% of the recombination energy contributes
in the ejection process ($\alpha_{\rm rec} = 0.5$).

Recent investigations have concluded that WD-MS binaries experience a strong orbital 
shrinkage during the CEP such that $\alpha$ should be $\sim 0.2-0.3$ without any additional
source of energy, like recombination energy, or at least with a very small contribution 
\citep[$\alpha_{\rm rec} \sim 0$),][]{Zorotovic_2010,Toonen_2013,Camacho_2014}.

With the aim of exploring the influence of more realistic values of the CEP parameters,
namely $\lambda$ and $\alpha$, we extend the set
of models studied by \citet{Belloni_2016a,Belloni_2016b}.
In this paper we simulate six new models, with exactly the same initial
conditions described in Table \ref{Tabmodels}, but with more reasonable
choices for the CEP parameters, namely: $\alpha=1$ and $\lambda=0.0$ (i.e. $\alpha_{\rm rec} = 0.0$). 
These assumed values imply that all of the variation in the binary orbital energy (of the
giant core and the MS star) contributes to the expulsion of the
CE, and that the binding energy parameter is variable and properly calculated assuming that
none of the recombination energy is used to assist the expulsion of the CE.
\citep[][Appendix A]{Claeys_2014}.

To summarize, we consider in this work the six models simulated by MOCCA
with the initial conditions given in Table \ref{Tabmodels}, but 
assuming either $\alpha=3$ and $\alpha_{\rm rec}=0.5$ 
or $\alpha=1$ and $\alpha_{\rm rec}=0.0$. Hereafter, we will indicate
the differences among the models K1, K2, K3, S1, S2, and S3, by including 
a subindex to indicate the assumed value of $\alpha$.  

The choice of exploring different assumptions for the common envelope 
ejection is new to this work, and will allow us to re-assess previous 
results \citep{Belloni_2016a,Belloni_2016b} and their sensitivity to our assumptions 
for the CEP parameters, in addition to any differences caused by different 
AML approaches (Sections \ref{cv_evolution} and \ref{CAML}).

After having described the main codes and assumptions involved in this work
as well as the 12 models analyzed, we turn our attention to our 
main results.

%%%%%%%%%%%%%%%%%%%%%%%%%%%%%%%%%%%%%%%%%%%%%%%%%%%%%%%%%%%%%%%%%%
% NEW SECTION
%%%%%%%%%%%%%%%%%%%%%%%%%%%%%%%%%%%%%%%%%%%%%%%%%%%%%%%%%%%%%%%%%%

\section{CLUSTER EVOLUTION AND CEP PARAMETERS}
\label{gc_evol_cep_par}

We initially calculated six models that differ mainly with respect to their IBPs, 
initial binary fractions and initial densities.
Additionally, for this work, we simulated the same initial models,
but with more realistic values for the CEP parameters. Hence,
we are able to compare the global evolution of our simulated clusters when 
each of these two different sets of CEP parameters is adopted.

All together, we have a total of 12 models. Among 
these models, we find cases with black hole subsystems that form, post-core
collapse clusters, intermediate-mass black holes and a cluster
with a relaxation time greater than a Hubble time (initially extremely sparse).
%NL How are initially sparse models "purely expanding"...I know the word "purely" is 
%my edit here, so hopefully that didn't confuse things.
%DB: I changed it.

PCEBs have shorter periods for lower values of $\alpha$, such that we 
find that the number of PCEBs produced in our simulated clusters is smaller for lower
values of $\alpha$, since many of them merge during the CEP. A more 
general conclusion is: {\it the lower is the value of $\alpha$, 
the longer is the initial minimum period of any PCEB progenitors} 
\citep[see][for more detail]{Zorotovic_2014b}. This is because any PCEBs 
that would have had an initial period shorter than the minimum period for PCEBs to merge 
during the CEP. Since the initial period is unchanged by the choice
of the CEP parameters, decreasing $\alpha$ causes the number of PCEBs to decrease, 
thus increasing the number of mergers during the CEP. 

As the majority of PCEBs in each model are WD-MS and WD-WD 
binaries (i.e. CV and AM CVn progenitors), which is a consequence of 
the adopted IMF, the number of WDs in the system is higher for clusters
simulated with $\alpha=1$ and $\alpha_{\rm rec}=0.0$.  During the 
merger process, the MS star is absorbed into the giant envelope 
and the outcome is a new giant. 
This giant will further evolve into a WD, with its type depending on the giant 
involved in the process. Similarly (albeit more complex), when
the secondary is a WD, the outcome of the coalescence process is a giant
that will likely evolve into a more massive WD \citep[for more details on 
the coalescence of common-envelope cores, see][Section 2.7.2]{Hurley_2002}.
Thus, instead of having the same amount of WD-MS and WD-WD binaries as in 
models evolved with $\alpha=3$ and $\alpha_{\rm rec}=0.5$, clusters evolved with 
$\alpha=1$ and $\alpha_{\rm rec}=0.0$ produce more single isolated WDs (i.e. less WD-MS
and WD-WD binaries).

Clusters simulated with $\alpha=1$ and 
$\alpha_{\rm rec}=0.0$ produce more WD-NS binaries. This is
easy to understand given how the formation of these objects
might be influenced by the CEP parameters. When two WD progenitors are
involved in two CEPs (one for each of the WDs), the final outcome of both
CEPs (WD-WD binary) has a smaller period for $\alpha=1$ and $\alpha_{\rm rec}=0.0$
than for $\alpha=3$ and $\alpha_{\rm rec}=0.5$ (provided, of course, the binary survives the CEP). 
In some cases, the WDs are so close that mass transfer
can proceed from a less massive He WD or C/O WD onto a more massive O/Ne/Mg WD. This can 
lead to the formation of a NS via accretion-induced collapse.
Indeed, if the more massive WD triggers electron-capture reactions at its centre
upon reaching the Chandrasekhar mass limit, then the burning which propagates 
outward makes central temperatures and pressures high enough to cause a
collapse of the WD, inducing a contraction of the remnant \citep[e.g.][]{Nomoto_1991}. 
These systems are interesting because they are potential progenitors of
low-mass X-ray binaries and millisecond pulsars in GCs. For example,
\citet{Rivera_2015} recently discovered near-ultraviolet counterparts to 
five millisecond pulsars in the GC 47 Tucanae, and all of them have
likely He WD companions, whose masses are $\sim$ $0.16 - 0.3$ M$_\odot$
with cooling ages that lie between $\sim$ $0.1 - 6$ Gyr. 
 
Despite the differences outlined above, the global cluster evolution is not affected by our 
choice for the CEP parameters. The time evolution of the main cluster properties remains 
about the same in all cases, including the rate of mass loss, the evolution of the core and half-mass
radii, the rates and types of dynamical interactions, the binding energies of the resulting binaries,
the binary fraction, etc. This result is expected since
most PCEBs are too hard to interact on sufficiently short timescales to provide a sufficient 
additional source of energy to change the cluster evolution \citep{Leigh_2016}.

%%%%%%%%%%%%%%%%%%%%%%%%%%%%%%%%%%%%%%%%%%%%%%%%%%%%%%%%%%%%%%%%%%
% NEW SECTION
%%%%%%%%%%%%%%%%%%%%%%%%%%%%%%%%%%%%%%%%%%%%%%%%%%%%%%%%%%%%%%%%%%
\section{WD-MS AND CV FORMATION RATES}
\label{fap_formation_rate}

In a previous paper \citep{Belloni_2016b}, we inferred that the formation
rate of WD-MS binaries that will later become present-day CVs in globular 
clusters is strongly dependent on WD formation (due to stellar evolution). This is 
even the case for those WD-MS binaries formed through dynamical interactions. Additionally,
given the time-delay between the WD-MS binary formation and CV formation,
we inferred that the overall CV formation rate in globular 
clusters could be nearly constant. However, we emphasized in that work that
this conclusion was based on a very small set of models.

In this section, we extend our previous analysis. Here, we bring to bear 
twice the number of models, each of which has more CVs relative to the old 
models. We again apply a two-sample Kolmogorov-Smirnov test to 
every pairwise combination of models. The alternative 
hypothesis of the test is that {\it the paired samples are not drawn from the 
same theoretical CV formation rate}, a null result that would 
support the conclusion of a model-independent CV formation rate.  That is, 
all CV populations form in a similar way, independent of the host cluster properties.

Upon comparing models with the highest numbers of CVs (S2$_{\alpha=1}$ 
and K2$_{\alpha=1}$), we reject the null hypothesis 
at the 99.9 per cent confidence level. This suggests that 
the CV formation rate differs from model to model. This result is 
based on much better statistics than in \citet{Belloni_2016b}.  Thus, 
in our new sample, we are able to rule out the
possibility that the CV formation rate in globular clusters 
is similar irrespective of the host cluster properties.

\begin{figure}
   \begin{center}
    \includegraphics[width=0.98\linewidth]{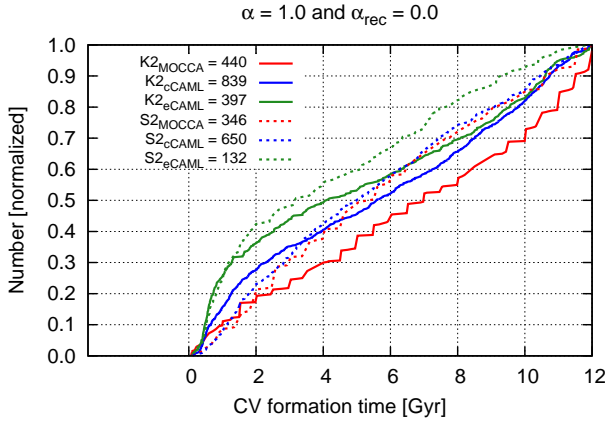} 
    \end{center}
  \caption{Cumulative distributions of CV formation times, normalized by the number of 
present-day CVs. As in previous plots, due to the high number of CVs in models S2$_{\alpha=1}$ and 
K2$_{\alpha=1}$, these models were used to make this plot. We consider here all three approaches 
for the AML/CV evolution (MOCCA, cCAML, and eCAML). In the figure keys, we indicate the number
of present-day CVs in each model/scheme.
For details, see Section \ref{fap_formation_rate}.}
  \label{Fig02}
\end{figure}

In Fig. \ref{Fig02}, we show the cumulative distributions 
for the formation-age CV populations, for models S2$_{\alpha=1}$ and K2$_{\alpha=1}$.  
We consider all three schemes for the AML/CV evolution 
(MOCCA, cCAML, and eCAML). Upon comparing the different models for
each AML/CV evolution scheme (same colour), we find that they are not the same.

As expected, for all models, when the eCAML formulation 
is adopted, a large fraction of CVs form early on in the
cluster evolution ($\lesssim$ 2 Gyr). This is associated with the number 
of massive WDs 
in the population (formed at $\lesssim$ 2 Gyr)\footnote{The cluster turn-off 
mass at low metallicity adopted in the simulations is $\sim$ 1.5 
M$_\odot$ at $\sim$ 2 Gyr. MS stars whose masses are smaller than this 
are likely to become He WDs, i.e. low-mass WDs, which are not present
using the eCAML scheme.} and the short 
time-delay between WD-MS binary formation and CV formation ($<1$ Gyr).
Additionally, most CVs that form early on in the cluster evolution belong
to the BSE group (i.e., formed without any influence from dynamics).

We emphasize that additional correlations between the CV formation
rate and the initial and present-day cluster properties might still exist.  
More models will be needed to address this fully.  
Although it is clear from our KS-tests that the CV formation rate differs from cluster to cluster,
CVs are continuously being added to the population throughout
the cluster evolution \citep{Belloni_2016b}. This is consistent with a nearly constant 
formation rate (in each cluster) after $\sim 1-2$ Gyr, 
with the ``constant of proportionality'' being different from cluster to cluster.
%NL: Hmm...Maybe say the "constant of proportionality" is different from cluster-to-cluster?  
%MZ: I agree. Linear coefficient also wasn't clear for me. Why not just ``slope''?
%DB: I think "constant of proportionality" seems okay.

Finally, we stress that looking for theoretical correlations between CV formation
rates and cluster properties is extremely important. For example, knowing this
and knowing the number of bright CVs observed in a particular GC would allow
us to account for the underlying observational biases and estimate the real number of CVs in 
such a GC, or even to use this information as a constraint for the initial GC conditions 
\citep[e.g.][]{Leigh_2013,Leigh_2015}.

%%%%%%%%%%%%%%%%%%%%%%%%%%%%%%%%%%%%%%%%%%%%%%%%%%%%%%%%%%%%%%%%%%
% NEW SECTION
%%%%%%%%%%%%%%%%%%%%%%%%%%%%%%%%%%%%%%%%%%%%%%%%%%%%%%%%%%%%%%%%%%

\section{NUMBER OF PRESENT-DAY CVS}
\label{number_pdp_cvs}

In Table \ref{Tab01}, we show the number of present-day CVs in all simulated clusters, 
separated according to the influence of dynamics for each formation channel. 
We emphasize that a 
few CVs formed adopting the MOCCA/cCAML/eCAML formulations for the WD-MS evolution do not 
survive in the cluster up to 12 Gyr. This is due to dynamical interactions or escape from 
the cluster or, mainly, mergers during unstable mass transfer.  For the numbers shown 
in Table \ref{Tab01}, only those CVs that form and survive to the present-day are included.
More details on those CVs that do not survive may be found in \citet[][section 3.8]{Belloni_2016b}.

In what follows, we discuss how CV production is affected by our choices for the CEP parameters
(Section \ref{number_pdp_cvs_CEP_par}), AML schemes (Section \ref{number_pdp_cvs_AML_sch}),
and initial cluster conditions (Section \ref{number_pdp_cvs_dynamics}).

%%%%%%%%%%%%%%%%%%%%%%%%%%%%%%%%%%%%%%%%%%%%%%%%%%%%%%%%%%%%%%%%%%
% NEW SECTION
%%%%%%%%%%%%%%%%%%%%%%%%%%%%%%%%%%%%%%%%%%%%%%%%%%%%%%%%%%%%%%%%%%

\subsection{Influence of CEP parameters}
\label{number_pdp_cvs_CEP_par}

\begin{table*}
\centering
\caption{Number of CVs formed over the course of the cluster evolution that 
survive to the present-day ($\sim$ 12 Gyr).  The CVs are separated according to 
the dynamical group (BSE, WDI, SDI) to which they belong, 
namely: the binary stellar evolution (BSE) group (without dynamics), the weak dynamical interaction 
(WDI) group and the strong dynamical interaction (SDI) group. The numbers correspond
to CVs produced purely in the MOCCA code, CVs produced assuming the classical version of 
CAML (cCAML), and CVs produced assuming the empirical formulation of CAML (eCAML). 
Models have either $\alpha=3$ and $\alpha_{\rm rec}=0.5$, 
or $\alpha=1$ and $\alpha_{\rm rec}=0$ (see Sections \ref{models} and \ref{number_pdp_cvs} for more details).
}
\label{Tab01}
\begin{adjustbox}{max width=\textwidth}
\noindent
\begin{threeparttable}
\noindent
\begin{tabular}{l|c|c|c|c|c|c|c|c|c|c|c|c|c|c|c|c|c|c|c|c}
\hline
Model & \multicolumn{5}{c}{MOCCA \tnote{a}} & & \multicolumn{6}{c}{cCAML \tnote{b}} & \multicolumn{5}{r}{eCAML \tnote{c}} \\
\hline
\specialcell{Influence of\\Dynamics} & BSE \tnote{d} & WDI \tnote{e} & \multicolumn{3}{c}{SDI \tnote{f}} & Total & & BSE  & WDI & \multicolumn{3}{c}{SDI} & Total & & BSE  & WDI & \multicolumn{3}{c}{SDI} & Total \\ 
\hline
\specialcell{Formation\\Channel} & CEP \tnote{g} & CEP & CEP & Exchange \tnote{h} & Merger \tnote{i} & & & CEP & CEP & CEP & Exchange & Merger & & & CEP & CEP & CEP & Exchange & Merger &  \\ 
\hline\hline	     % CEP   CEP  CEP   EXC  MER   TOTAL     CEP   CEP  CEP   EXC  MER    TOT	   CEP   CEP  CEP  EXC  MER   TOT	
K$1_{\rm \alpha=3}$ &   0 &   0 &  1 &   2 &  0 &    3  & &   0 &   0 &   3 &   2 &  0 &    5 & &   0 &   0 &   0 &  0 &  0 &   0 \\ \hline
K$2_{\rm \alpha=3}$ &   0 &  19 & 42 & 106 & 15 &  182  & &   0 &  38 &  62 & 151 & 36 &  287 & &   0 &  20 &  33 & 76 & 15 & 144 \\ \hline
K$3_{\rm \alpha=3}$ &   0 &   9 & 22 &  71 & 28 &  130  & &   0 &  27 &  47 &  96 & 46 &  216 & &   0 &  21 &  35 & 54 & 28 & 138 \\ \hline
S$1_{\rm \alpha=3}$ &  40 &   1 &  0 &   1 &  0 &   42  & &  81 &   1 &   1 &   1 &  0 &   84 & &  20 &   0 &   0 &  1 &  0 &  21 \\ \hline
S$2_{\rm \alpha=3}$ & 122 & 107 & 17 &  11 &  3 &  260  & & 224 & 178 &  29 &  30 &  4 &  465 & &  42 &  29 &  12 & 10 &  1 &  94 \\ \hline
S$3_{\rm \alpha=3}$ &   9 &  14 &  2 &   0 &  2 &   27  & &  10 &  21 &   3 &   4 &  1 &   39 & &   3 &   1 &   0 &  1 &  0 &   5 \\ \hline
\hline\hline
K$1_{\rm \alpha=1}$ &  19 &  19 &  3 &   2 &  0 &   43  & & 101 & 109 &   9 &   5 &   0 &  224 & &  74 &  65 &   0 &   1 &  0 & 140 \\ \hline
K$2_{\rm \alpha=1}$ &   9 &  80 & 77 & 209 & 65 &  440  & &  64 & 260 & 154 & 254 & 107 &  839 & &  46 & 143 &  40 & 121 & 47 & 397 \\ \hline
K$3_{\rm \alpha=1}$ &   4 &  36 & 30 &  76 & 18 &  164  & &  25 & 140 &  52 &  96 &  54 &  367 & &  14 & 104 &  35 &  46 & 27 & 226 \\ \hline
S$1_{\rm \alpha=1}$ &  76 &   2 &  0 &   0 &  0 &   78  & & 148 &   3 &   0 &   1 &   0 &  152 & &  26 &   1 &   0 &   0 &  0 &  27 \\ \hline
S$2_{\rm \alpha=1}$ & 139 & 123 & 13 &  52 & 19 &  346  & & 281 & 240 &  25 &  62 &  42 &  650 & &  41 &  40 &   4 &  29 & 18 & 132 \\ \hline
S$3_{\rm \alpha=1}$ &   7 &   4 &  0 &   3 &  2 &   16  & &  21 &   6 &   0 &   3 &   5 &   35 & &   1 &   1 &   0 &   1 &  2 &   5 \\ \hline
\hline 
\end{tabular}
\begin{tablenotes}
       \item[] $^{\rm a}$ CVs formed in the MOCCA code (no CAML); 
$^{\rm b}$ CVs formed when the classical CAML is assumed; 
%NL: The one below is the same as the one above, isn't it?  Is this a typo?
%DB: In fact. :) Monica corrected.
$^{\rm c}$ CVs formed when the empirical CAML is assumed; 
$^{\rm d}$ CVs that are formed without any influence from dynamics; 
$^{\rm e}$ CVs that are formed with only a weak influence from dynamics; 
$^{\rm f}$ CVs that are formed with a strong influence from dynamics; 
$^{\rm g}$ CVs formed via a CEP; 
$^{\rm h}$ CVs formed via a dynamical exchange;
$^{\rm i}$ CVs formed via a dynamical merger;   
\end{tablenotes}
\end{threeparttable}
\end{adjustbox}
\end{table*}

In Section \ref{gc_evol_cep_par}, we found that {\it the number of PCEBs decreases 
when $\alpha$ and $\alpha_{\rm rec}$ decrease}.
Even though the number of PCEBs is reduced in our 
models with $\alpha = 1$ and $\alpha_{\rm rec} = 0.0$,
this affects the number of CVs in the opposite way. 
In other words, we do not see a reduction in the number of CVs caused by the reduction
in the number of PCEBs that are WD-MS binaries.  Instead, the opposite occurs,
i.e. the total number of CVs produced is larger for $\alpha=1$ and $\alpha_{\rm rec}=0.0$ 
in all models but model S3 (see Section \ref{number_pdp_cvs_dynamics}). 

This can be understood by looking at the final configuration of
PCEBs. Lower values for the CEP parameters lead to shorter periods.  
This causes PCEBs to remain in the PCEB phase for shorter durations.  In turn, 
this produces more CVs since the pre-CV lifetime is reduced and more systems become CVs 
in less than 12 Gyrs.
%NL:  I am not sure I understand the paragraph above.  Are you saying that the shorter 
%periods make it less likely that a dynamical interaction will destroy the CV before 12 Gyr?
%DB: No. The idea is that more WD-MS become CVs. If the CEP parameters are smaller, 
%the final orbital period is shorter (for the same ZAMS binary). Within 10-12 Gyr,
%more WD-MS will become CVs, for lower values of the CEP paramters. This is not connected
%with dynamics, a priori.

Regardless of the assumed AML and CV evolution prescriptions, there are no 
CVs produced in the Kroupa models when $\alpha=3$ and $\alpha_{\rm rec}=0.5$ are
adopted. Only by decreasing $\alpha$ and $\alpha_{\rm rec}$ 
can the Kroupa models form CVs without any influence from dynamics. As suggested 
by previous works based on adopting the Standard IBP, the assumed 
CEP parameters should be small. This seems to be the case for the
Kroupa IBP as well.

%%%%%%%%%%%%%%%%%%%%%%%%%%%%%%%%%%%%%%%%%%%%%%%%%%%%%%%%%%%%%%%%%%
% NEW SECTION
%%%%%%%%%%%%%%%%%%%%%%%%%%%%%%%%%%%%%%%%%%%%%%%%%%%%%%%%%%%%%%%%%%

\subsection{Influence of AML schemes}
\label{number_pdp_cvs_AML_sch}

With respect to the CV evolution scheme, the number of CVs formed in MOCCA 
is smaller than the number of CVs formed adopting the cCAML formulation. This
is associated with the stability criterion for dynamical mass transfer. In MOCCA,
the criterion for a CV to start unstable mass transfer in a dynamical timescale
is fixed ($q > q_0 = 0.695$). In the case of cCAML, this critical mass ratio depends
on the mass of the secondary and it can be as great as $\sim 1.2$. Thus, many CVs 
that survive when cCAML is adopted merge in the pure MOCCA simulations. In the case
of eCAML, the critical mass ratio also depends on the secondary mass. For low-mass 
secondaries it is more strict than the fixed value from MOCCA and also than the 
variable limit for cCAML, but it can reach extremely large values for more massive
companions \citep{Schreiber_2016}.

Interestingly, the relative numbers of CVs produced adopting the eCAML formulation 
is greater in the Kroupa models, compared to adopting the cCAML formulation. This indicates 
that the Kroupa IBP produces more C/O WDs (high-mass) than the Standard IBP, especially 
in model K$1_{\rm \alpha=1}$. This cluster is extremely sparse initially 
(i.e., only a few CVs form dynamically). Even so, $\sim$ 60 per cent of the CVs formed with the 
cCAML scheme are also present when eCAML is assumed. The same is not true for model S$1_{\rm \alpha=1}$, 
which is sparse as well. In this cluster, only $\sim$ 17 per cent of the CVs formed with the cCAML 
approach are present with the eCAML approach.

More generally, as expected, the fraction of CVs that remain with the eCAML formulation is 
around 15-20 per cent for the S models (Standard IBP), as pointed out by \citet{Schreiber_2016}. 
On the other hand, for the K models (Kroupa IBP), this fraction is $\sim$ 50-60 per cent. 
As mentioned above, this indicates that upon assuming a Kroupa IBP the production of CVs with high-mass WDs
is more efficient, regardless of the formation channel.

%%%%%%%%%%%%%%%%%%%%%%%%%%%%%%%%%%%%%%%%%%%%%%%%%%%%%%%%%%%%%%%%%%
% NEW SECTION
%%%%%%%%%%%%%%%%%%%%%%%%%%%%%%%%%%%%%%%%%%%%%%%%%%%%%%%%%%%%%%%%%%

\subsection{Influence of the initial cluster conditions}
\label{number_pdp_cvs_dynamics}

In general, for both IBPs, {\it the denser the cluster is
initially, the smaller the number of CVs formed through stellar evolution alone}.
In fact, upon comparing the columns associated with the BSE group in Table
\ref{Tab01}, for all schemes (MOCCA, CAML and eCAML), we see a reduction
in the relative number of CVs formed in the BSE group.  
This indicates that, for all clusters, {\it any reduction in the 
relative numbers of CV progenitors is correlated with the
initial cluster density}, which is associated with the
role of dynamical interactions in destroying CV progenitors,
which is in turn related to the cluster soft-hard boundary.

The semi-major axis that defines the boundary between soft and hard binaries
is proportional to $r_h/N$, where $r_h$ is the half-mass radius and $N$ is the
number of objects (single + binaries) in the cluster. In addition, for clusters
with similar $N$ (as is the case in the three Kroupa models), the denser the cluster
(i.e. the smaller the half-mass radius), the smaller the semi-major axis that 
defines the soft-hard boundary. Therefore, the soft-hard boundary is inversely
related to the cluster density, and as the cluster density increases, the 
soft-hard boundary also decreases (in semi-major axis/period).
At a particular density, the soft-hard boundary will enter in to the region occupied by 
CV progenitors.  Thus, beyond this density, more and more CV
progenitors are potentially destroyed, as the density increases. This is the
reason behind the above-mentioned correlation between the cluster density
and the number of CV progenitors that become CVs via binary stellar evolution.

As already mentioned, the number of CVs in model S3 is smaller for lower values
of the CEP parameters. This cluster is very dense and dynamics plays 
a huge role, due to the short orbital separation corresponding to the soft-hard 
boundary. For the lower values of $\alpha$ and $\alpha_{\rm rec}$, the number of WD-MS binaries 
in the cluster is about half the analogous numbers when the cluster is evolved 
with the larger CEP parameters. This is not the case for the S1 and S2 models where dynamical effects are not as strong. 
In these two models, for the lower CEP efficiencies, around 80\% of WD-MS binaries 
survive the CEP in comparison to the number of WD-MS binaries obtained with the higher CEP efficiencies. 
In model S3, we adopt the smaller CEP efficiencies, such that a large number of CV progenitors come from initial binaries whose orbital
separations are greater than the orbital separation that defines the soft-hard 
boundary. For these `wide' binaries, the probability of interaction is larger and 
they can be more easily destroyed by dynamics. 
%NL:  The above paragraph could benefit from some slight re-wording for clarity, since I did not 
%always follow the arguments.  I did my best up until the half-way point anyways.
%DB: I changed it. Could you please check? I see now that this is hard to explain. :(
%What I tried to say is the in S3, the soft-hard boundary enters deeper in the
%CV progenitor semi-major axis distribuiton, which causes a lot of CV progenitor 
%destruction.
%NL: Gotcha.  I made some edits, but only for language.

Most CVs in the Standard models 
are formed via a CEP, which may be preceded 
by weak dynamical interactions. Usually, strong dynamical interactions
do not play a significant role, which is directly
related to the fact the Standard IBP contains more
hard binaries than the Kroupa IBP (see appendix \ref{ap}). However, 
the influence of strong dynamical interactions increases slightly for 
the eCAML formulation. This is easy to understand based on the
role of exchanges in the CV formation process, which puts more massive 
WDs into the CV population \citep[e.g.][]{Belloni_2016b}. Thus, the probability
of surviving the stability criterion imposed by the eCAML scheme is enhanced
for CVs formed via strong dynamical interactions.

As already pointed out by \citet{Belloni_2016a,Belloni_2016b},
when $\alpha=3$ and $\alpha_{\rm rec}=0.5$, models that follow the Kroupa IBP produce 
only dynamically formed CVs, with exchanges being the main formation channel. Here we 
confirm that the same is obtained when CAML is included. 
When lower values are adopted for the CEP parameters ($\alpha=1$ and 
$\alpha_{\rm rec}=0.0$), the Kroupa models also have CVs formed through
only stellar evolution (BSE group). 
Although field-like CVs can form in all Kroupa models with lower CEP efficiencies, only
in the sparsest cluster (K1$_{\alpha=1}$) do they contribute significantly 
to the overall CV population. This is related to the fact that
dynamically formed CVs are rare in sparse clusters.

%%%%%%%%%%%%%%%%%%%%%%%%%%%%%%%%%%%%%%%%%%%%%%%%%%%%%%%%%%%%%%%%%%
% NEW SECTION
%%%%%%%%%%%%%%%%%%%%%%%%%%%%%%%%%%%%%%%%%%%%%%%%%%%%%%%%%%%%%%%%%%
\section{PRESENT-DAY CV PROPERTIES}
\label{pdp}

In this section, for convenience,
we show only the results for models K2 and S2, for both
choices of the CEP parameters. This is because these two models produce the largest 
numbers of CVs, which increases the statistical significance
of our analysis. This is a reasonable approach since models with the 
same IBP tend to have similar properties in the end. However, 
the relative numbers of CVs formed from each channel depends on the 
initial cluster conditions, as we saw in Section \ref{number_pdp_cvs_dynamics}.

In Sections \ref{pdp_secondary_mass} and \ref{pdp_period}, we discuss the CV 
component massses and periods. Subsequently, we address the 
mass transfer rate (Section \ref{pdp_mtr}), duty cycle (Section \ref{pdp_dc}), 
X-ray luminosity (Section \ref{pdp_lx}), quiescent magnitudes (Section \ref{pdp_mag})
and spatial distribution (Section \ref{pdp_spatial}).

%%%%%%%%%%%%%%%%%%%%%%%%%%%%%%%%%%%%%%%%%%%%%%%%%%%%%%%%%%%%%%%%%%
% NEW SECTION
%%%%%%%%%%%%%%%%%%%%%%%%%%%%%%%%%%%%%%%%%%%%%%%%%%%%%%%%%%%%%%%%%%
\subsection{WD mass}
\label{pdp_secondary_mass}

The main difference in the WD mass distribution 
(left-hand column of Fig. \ref{Fig04.1}) for the different
AML formulations is the absence of low-mass WDs when eCAML is assumed.
As already stated in Section \ref{CAML}, the empirical formulation 
for CAML is motivated by the absence of CVs with He WD primaries
\citep{Schreiber_2016}. Consequently, it is not a surprise
that the WD mass distributions that use the eCAML scheme contain
only C/O and O/Ne/Mg WDs (i.e. massive WDs). The minimum
WD mass in the distribution for eCAML is $\sim$ 0.6 M$_\odot$
(with a median value of $\sim$ 0.73 M$_\odot$) independent of the host 
cluster properties, CEP parameters, formation channel or 
IBP. This is illustrated in Fig. \ref{Fig04.1} (left-hand column)
which shows the WD mass distribution for models K2 and S2,
at the moment of CV formation, i.e. at the onset of mass transfer.
For the MOCCA and cCAML schemes, on the other hand, the WD mass can 
reach values as low as $\sim$ 0.2 M$_\odot$.

The WD mass distribution is also affected by the IBP and CEP 
parameters. For the Kroupa models, the distributions have a 
strong peak at $\sim$ 0.7 M$_\odot$. 
In the Standard models, another peak is also present at lower masses 
($\sim$ 0.3 M$_\odot$) which is absent in the eCAML approach.
The peak at $\sim$ 0.7 M$_\odot$ broadens and moves towards lower 
values ($\sim$ 0.6 M$_\odot$) when smaller efficiencies for the 
CEP are assumed. 
The interquartile range of the WD mass distribution
for the Standard IBP models (including all formation 
channels) is $[\sim 0.35; \sim 0.65]$.  This indicates that the majority 
of the CVs formed in these models have low-mass WDs. On the other
hand, upon considering the Kroupa IBP models, again
for all formation channels, the interquartile range is $[\sim 0.61; \sim 0.78]$.  
This indicates that most CVs formed in these models have high-mass
WDs. 
This is the main reason why the Kroupa models are less affected by the 
eCAML approach than are the Standard models.
The differences in the interquartile range are due to the intrinsic
differences between both IBPs adopted here in the region of parameter space 
corresponding to the CV progenitors (see Appendix \ref{ap}). 
The Kroupa IBP produces CVs with properties that more closely resemble 
those CVs observed in the Galactic field.  But, with that said, a more thorough
investigation that considers a realistic Galactic star-formation rate and observational 
selection effects is still to come.

CVs strongly influenced by dynamics (SDI group),
mainly have massive WDs ($\sim$ 75 per cent of the
CVs in the SDI group have masses greater than $\sim$ 0.6 M$_\odot$).
This is due to dynamical exchanges in which a binary composed of 
low-mass MS stars has one of its components replaced by either a more massive 
MS star or a more massive WD \citep{Belloni_2016b}. In both cases, the resulting 
CV WD mass is higher than it would otherwise have been without the exchange.

%%%%%%%%%%%%%%%%%%%%%%%%%%%%%%%%%%%%%%%%%%%%%%%%%%%%%%%%%%%%%%%%%%
% NEW SECTION
%%%%%%%%%%%%%%%%%%%%%%%%%%%%%%%%%%%%%%%%%%%%%%%%%%%%%%%%%%%%%%%%%%
\subsection{Donor mass and period}% and eccentricity}
\label{pdp_period}

\begin{figure*}
   \begin{center}
    \includegraphics[width=0.39\linewidth]{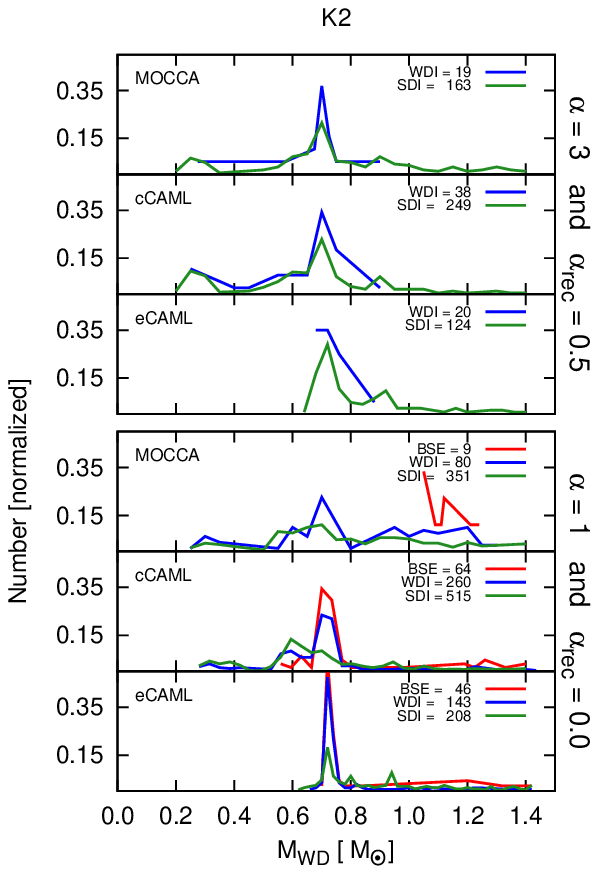}
    \hspace{1cm}
    \includegraphics[width=0.39\linewidth]{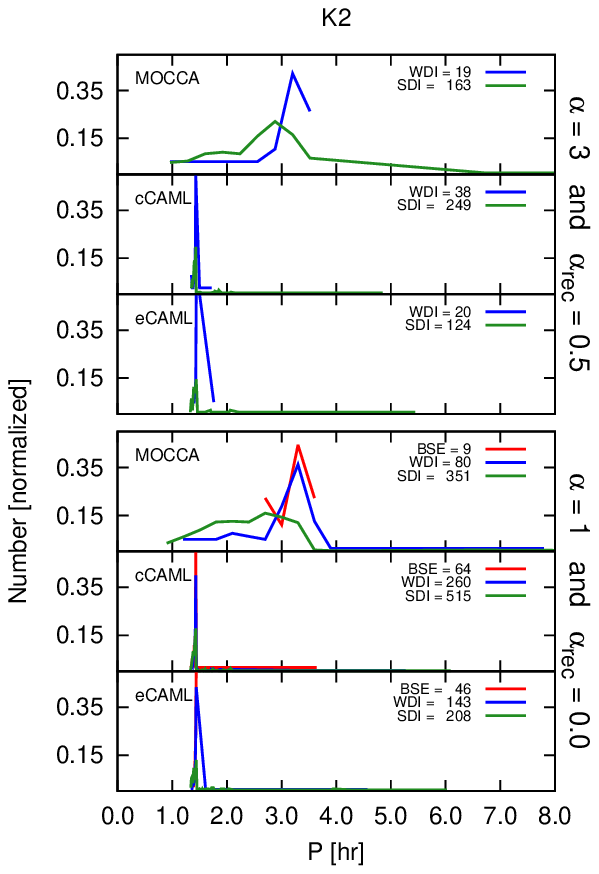}  
    \includegraphics[width=0.39\linewidth]{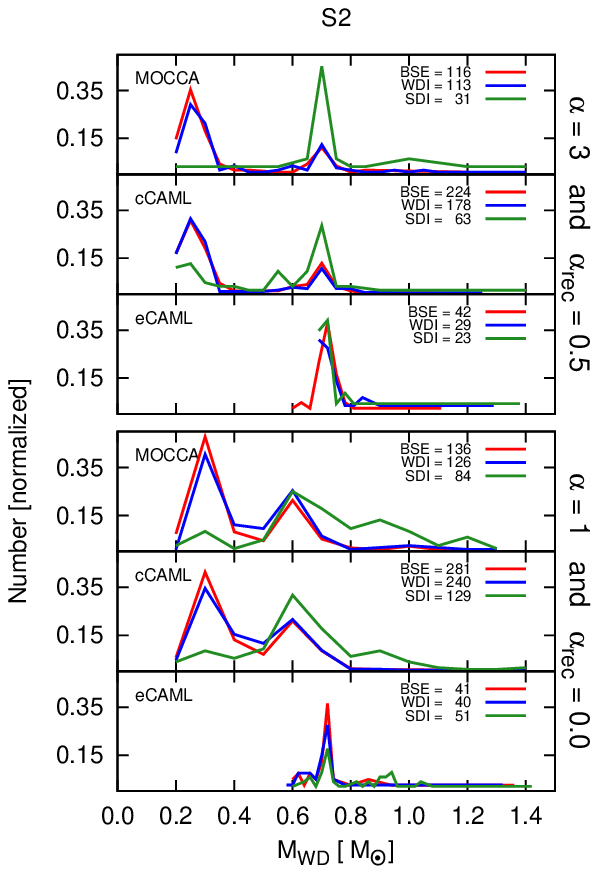}
    \hspace{1cm}
    \includegraphics[width=0.39\linewidth]{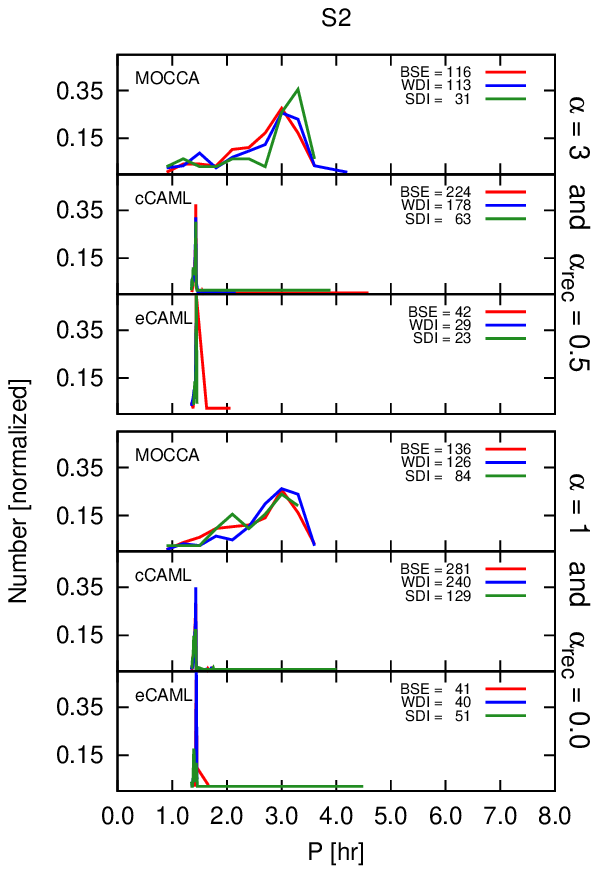}  
    \end{center}
  \caption{WD mass (left-hand column) and period (right-hand column)
distributions of present-day CVs, normalized with respect to the total 
number of objects in each dynamical group (BSE, WDI, and SDI). 
In the upper row, we show the  distributions for model K2 (Kroupa IBP) and in the
bottom row, for model S2 (Standard IBP). The first three rows in each panel
correspond to $\alpha=3$ and $\alpha=0.5$, and the next three rows correspond 
to $\alpha=1$ and $\alpha=0.0$. The keys indicate the number of CVs in 
each dynamical group (BSE, WDI, and SDI) for the three approaches considered here 
(MOCCA, cCAML, and eCAML). For more details, see Sections \ref{pdp_secondary_mass} 
and \ref{pdp_period}.}
  \label{Fig04.1}
\end{figure*}

\begin{figure*}
   \begin{center}
    \includegraphics[width=0.39\linewidth]{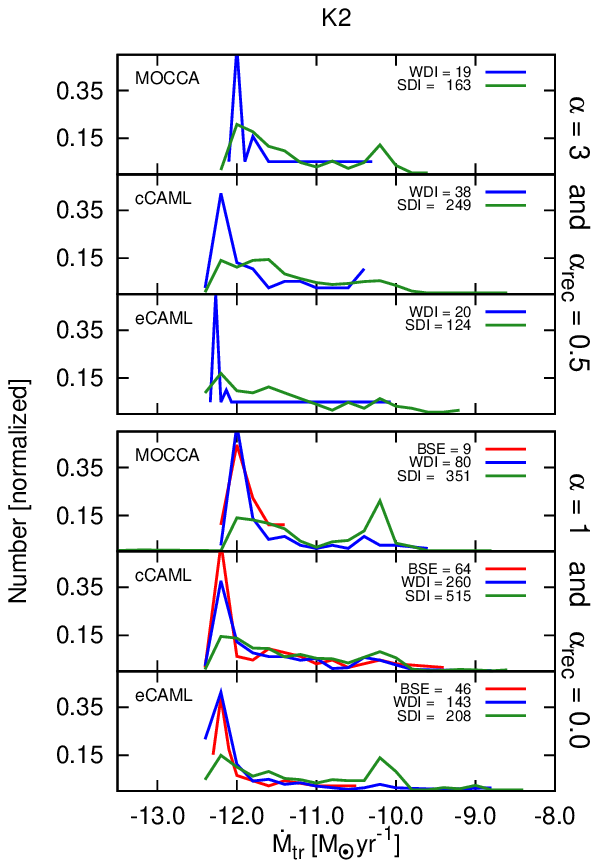} 
    \hspace{1cm}
    \includegraphics[width=0.39\linewidth]{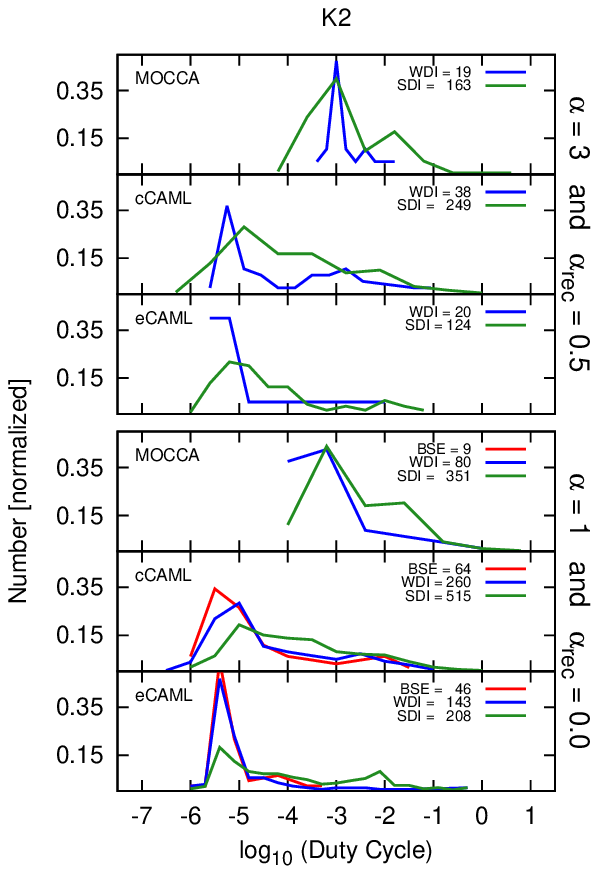} 
    \includegraphics[width=0.39\linewidth]{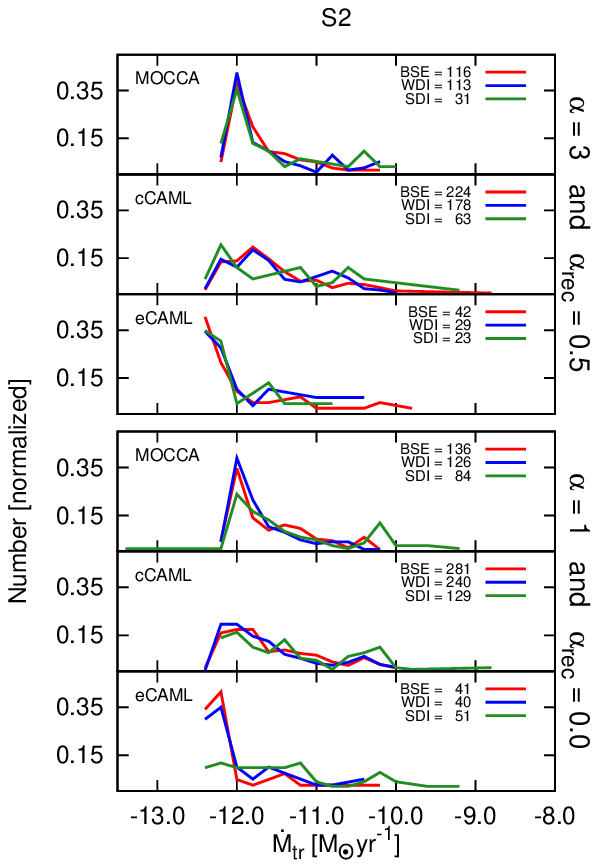} 
    \hspace{1cm}
    \includegraphics[width=0.39\linewidth]{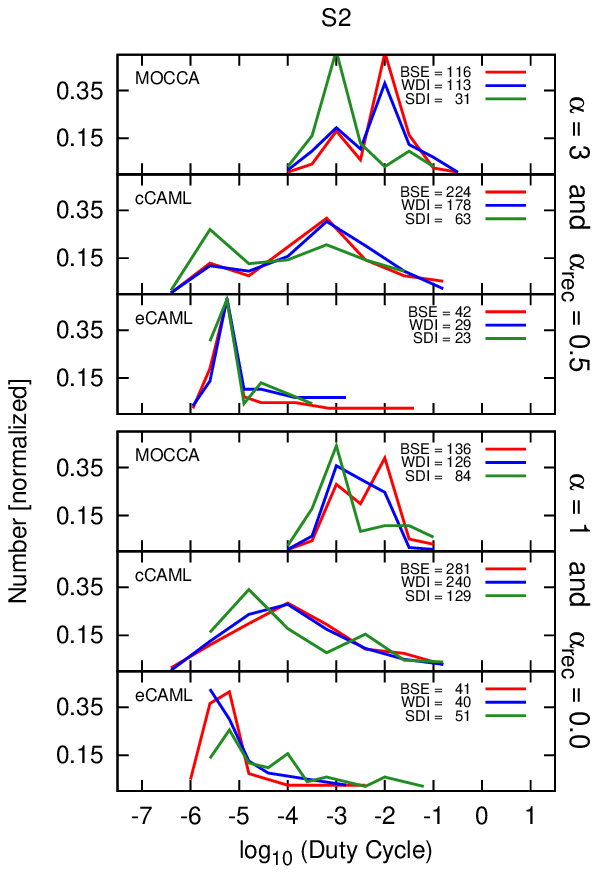} 
    \end{center}
  \caption{Mass transfer rate (left-hand columns) and duty cycle (right-hand column) 
distributions of present-day CVs, normalized with respect to the total 
number of objects in each dynamical group (BSE, WDI, and SDI). 
In the upper row, we show the  distributions for model K2 (Kroupa IBP) and in the
bottom row, for model S2 (Standard IBP).
The layout of this figure is the same as in Fig. \ref{Fig04.1}. 
For more details, see Sections \ref{pdp_mtr} and \ref{pdp_dc}.}
  \label{Fig04.2}
\end{figure*}

For the donor masses, we find an accumulation
of systems at the very low-mass end (below the brown dwarf limit)
in all the simulations, regardless of the IBP, AML prescription 
or CEP parameters. CVs simulated with MOCCA have in general
higher secondary masses than with cCAML/eCAML. This is mainly
associated with the donor mass-radius relation adopted in each
code (see Section \ref{discussion_CVs_aml}).
%MZ This is so strange. They start in MOCCA with smaller M2's. They
%also have higher mass transfer rates. Then they should evolve faster
%towards smaller masses. Might not be that BSE just stops below some value
%for M2?
%DB: BSE evolves the CVs for an indefinite time, I think. At least the
%CV remains as CV while increasing the period. When I simply changed
%the mass-radius relation (I used the Knigge+2011 one), the CV evolved
%as in the Knigge+2011 tracks. Similar to your code. I think that having
%a mass-radius relation for BDs similar to that for WDs causes a reduction
%in the mass loss and an enhanced period variation. At least, comparing
%both mass-radius relations, that is the impression I got.
Therefore, most CVs are period-bouncers, irrespective of the cluster 
properties, CEP parameters, formation channel or IBP. 
In general, CVs formed with a strong influence from dynamics 
(SDI group) have a lower fraction of period-bouncers,
which implies that they have a larger probability of being observed during
quiescence (due to higher mass transfer rates). However, for the Standard
models, when considering either the cCAML or eCAML approaches with 
$\alpha=3$ and $\alpha_{\rm rec}=0.5$, the opposite occurs, i.e., for these 
models, the fraction of period-bouncers is
higher in the set of dynamically formed CVs. This is because
most CVs in the SDI group are formed at earlier times.

The period distributions are shown 
in the right-hand column of Fig. \ref{Fig04.1}. Here we see a drastic
difference in the CV evolution model implemented
by MOCCA relative to the state-of-the-art model implemented in the cCAML/eCAML
scheme. This is mainly due to the mass-radius relation adopted in BSE 
\citep[][see Eq. 24]{Hurley_2000}, that assumes a substantial increase 
in the radius of the donor when its mass decreases, for secondary masses 
below $\sim$ 0.1 M$_\odot$. This results in period bouncers that can 
reach periods longer than 3 h, which is not realistic \citep{Knigge_2011_OK}. 
%NL By "realistic", do you mean that it is not seen in the observed CV populations?  If so, I would say it more explicitly.
%DB: It is hard to say that this is related to observations, due to biases. This would be with respect to models, that can
%reproduce observed properties of CVs in the solar vicinity. I added a citation to Knigge. Do you think it is okay?
%For period bouncers, everything is very uncertain, actually.

On the other hand, in the state-of-the-art CV evolution model, the mass-radius relation
adopted \citep{Knigge_2011_OK} assumes that below $\sim$ 0.07 M$_\odot$
the radius decreases more slowly as the mass decreases, and at no point does it 
increase. Therefore, period bouncers remain close to the period 
minimum for longer. However, the authors caution that this mass-radius 
relation should not be extrapolated below $\sim$ 0.05 M$_\odot$, although
the systems should remain close to the period minimum and with very low 
mass transfer rates (probably undetectable). 

%%%%%%%%%%%%%%%%%%%%%%%%%%%%%%%%%%%%%%%%%%%%%%%%%%%%%%%%%%%%%%%%%%
% NEW SECTION
%%%%%%%%%%%%%%%%%%%%%%%%%%%%%%%%%%%%%%%%%%%%%%%%%%%%%%%%%%%%%%%%%%
\subsection{Mass transfer rate}
\label{pdp_mtr}

For the mass transfer rate distribution (left-hand column
of Fig. \ref{Fig04.2}), the main differences between models are associated with
the CV evolution scheme and the different formation channels. 

Although all schemes are concentrated towards low mass transfer rates,
the peak in the distribution for systems evolved with the MOCCA scheme
is at higher values than for those that incorporate CAML. 
%DB: I replaced "models"/"code" with "schemes"/"those".
As already discussed in Sect. \ref{pdp_period}, this is related to the
mass radius relation adopted in the BSE code used in MOCCA as well 
as an inadequate mass transfer rate equation for the AML timescale.
%DB: I also added info about wrong MTR equation for MTR driven by AML.
%Please check if it is okay.
%NL: I think it seems okay, but did some very very slight re-wording for language.
%DB: Okay. 

For CVs formed without any or with only a weak 
influence from dynamics (BSE and WDI groups), most CVs have mass transfer
rates smaller than $\sim 5 \times 10^{-12}$ M$_\odot$/yr, especially in
the Kroupa models, for both sets of CEP parameters. For
those CVs strongly influence by dynamics (SDI group), 
the distribution is flatter.
This indicates that CVs in the SDI group should be brighter, in general,
than CVs in either the BSE or WDI groups. As we will see in Section \ref{pdp_mag},
this is in fact the case.

Applying the DNe criterion, as described in \citet[][see Section 3.3.1]{Belloni_2016a}, we can
separate the CVs into three groups with respect to the instability of their disks,
namely CVs whose disks are hot/stable, unstable, or cold/stable.

For most CVs in our models, the disks are unstable, with very
few exceptions of stable disks (less than 1 per cent). 
Additionally, most of the disks exhibit outbursts of Type B (i.e. inside-out)
that begin in the inner parts of the disk and propagate outward
\citep[e.g.][]{Smak_2001,Lasota_2001}.  This is due to their extremely small
mass transfer rates. Thus, if such CVs do not have WDs with relatively 
high magnetic fields such that the inner part of their 
disks might be disrupted \citep{Dobrotka_2006}, 
practically all CVs in our models are DNe and undergo outbursts.
Most GC CVs seem to be DNe \citep[e.g.][]{Servillat_2011,Knigge_2011_OK,Knigge_2012MMSAI,Webb_2013,Belloni_2016a}.  
As we will see in what 
follows, the occurrence of outbursts for such CVs is rather infrequent, which 
makes their detection through outbursts very difficult.
%NL: I thought a reference would be useful where I indicated it above, to help support the point.
%DB: I added some references.

%%%%%%%%%%%%%%%%%%%%%%%%%%%%%%%%%%%%%%%%%%%%%%%%%%%%%%%%%%%%%%%%%%
% NEW SECTION
%%%%%%%%%%%%%%%%%%%%%%%%%%%%%%%%%%%%%%%%%%%%%%%%%%%%%%%%%%%%%%%%%%
\subsection{Duty cycle}
\label{pdp_dc}

One important property of DN CVs with regard to their detectability is
the duty cycle. This can be defined as the fraction of the DN cycle 
in which the DN is in outburst, i.e. 
the ratio between the duration of the outburst and the
recurrence time. In order to compute these two quantities,
we use empirical relations as described in 
\citet[][see Section 3.3.5]{Belloni_2016a}.

%%%%%%%%%%%%%%%%%%%%%%%%%%%%%%%%%%%%%%%%%%%%%%%%%%%%%%%%%%%%%%%%%%
% NEW SECTION
%%%%%%%%%%%%%%%%%%%%%%%%%%%%%%%%%%%%%%%%%%%%%%%%%%%%%%%%%%%%%%%%%%
\subsubsection{Recurrence time}
\label{pdp_dc_rt}

Due to the differences in the donor mass and period distributions
previously discussed, the CV recurrence times with the MOCCA
formulation are smaller than those for CVs formed with the cCAML and
eCAML formulations. This is because the recurrence time is inversely 
proportional to the mass ratio, and the mass ratios for MOCCA CVs 
are greater. Specifically, most CVs formed in MOCCA have recurrence
times shorter than $10^4$ days, for both sets of CEP parameters.

With the cCAML formulation, we see a clear difference
with respect to the IBP. When $\alpha=3$ and $\alpha_{\rm rec}=0.5$
are adopted, the median of the distribution is $\sim 10^4$ days 
for the Kroupa models and $\sim 10^3$ days for the Standard models. 
When $\alpha=1$ and $\alpha_{\rm rec}=0.0$ are adopted,
the median of the distribution for the Kroupa models is similar, while 
for the Standard models the median increases and becomes similar
to what is seen in the Kroupa models.

With the eCAML formulation, regardless of the assumed CEP parameters
and the IBP, most CVs have recurrence times longer than $\sim 10^5$ days.
They are longer here because the mass ratios are much smaller, since 
these CVs are dominated by massive WDs.

In general, CVs strongly influenced by dynamics (SDI group) have longer 
recurrence times. This is because, usually, they are formed from more 
massive WDs.

%%%%%%%%%%%%%%%%%%%%%%%%%%%%%%%%%%%%%%%%%%%%%%%%%%%%%%%%%%%%%%%%%%
% NEW SECTION
%%%%%%%%%%%%%%%%%%%%%%%%%%%%%%%%%%%%%%%%%%%%%%%%%%%%%%%%%%%%%%%%%%
\subsubsection{Outburst duration}
\label{pdp_dc_od}

For the duration of the outburst, which is associated with the
period (the longer the period, the greater is the extent of the disk and the longer
is the outburst), we see a clear distinction upon comparing the 
MOCCA and cCAML/eCAML formulations. As noted before, most CVs formed in the 
cCAML and eCAML simulations have similar periods, which is close
to the period minimum (in turn, the outbursts have similar durations). 
On the other hand, CVs in MOCCA simulations have much longer periods
(up to $\sim 3-4$ h), which makes the outbursts last longer. Even
so, the duration of the outbursts is just a few days. This is the case 
for all models, CEP parameters and AML/CV evolution schemes.

We emphasize here that the empirical relation used to compute
the duration of the outbursts was derived from well-observed
CVs in the solar neighbourhood \citep{Smak_1999}.
It includes only the normal outburst. This is, apparently, not the
case for period-bouncers, where superoutbursts play a more important
role and could increase the duration time significantly, since the outbursts 
can last for months. We discuss how this affects our results in more 
detail in Section \ref{discussion_CVs_emp}.

%%%%%%%%%%%%%%%%%%%%%%%%%%%%%%%%%%%%%%%%%%%%%%%%%%%%%%%%%%%%%%%%%%
% NEW SECTION
%%%%%%%%%%%%%%%%%%%%%%%%%%%%%%%%%%%%%%%%%%%%%%%%%%%%%%%%%%%%%%%%%%
\subsubsection{Duty cycle}
\label{pdp_dc_dc}

Finally, we present our main results regarding the behaviour of two 
important timescales associated with DNe. Thereafter, we will turn our 
attention to a description of the duty cycle distribution, which
is illustrated in the right-hand column of Fig. \ref{Fig04.2}.

Since the duration time is similar for all models and formulations,
the main difference in the duty cycle comes from the recurrence
time. 
The duty cycles of CVs formed in the MOCCA simulations are usually
larger than $\sim$ 0.1 per cent (first quartile), and smaller 
than $\sim$ 1 per cent (third quartile), for both choices of
CEP parameters.
On the other hand, for the cCAML and eCAML formulations, most CVs
have duty cycles smaller than $\sim$ 0.1 per cent and $\sim$
0.001 per cent, respectively. This implies that GC CVs are 
even more difficult to detect upon adopting more realistic and
accurate CV evolution schemes. This corroborates our previous
result \citep{Belloni_2016a}, where we concluded the same thing based on the 
results of pure MOCCA simulations, which are out-dated with respect to their CV
evolution schemes.

As for the recurrence times, CVs strongly influenced by dynamics usually have
smaller duty cycles. This is because they usually have more massive WDs and
smaller mass ratios. However, we caution that those CVs in the SDI group that have 
low-mass WDs can still have large duty cycles.

%%%%%%%%%%%%%%%%%%%%%%%%%%%%%%%%%%%%%%%%%%%%%%%%%%%%%%%%%%%%%%%%%%
% NEW SECTION
%%%%%%%%%%%%%%%%%%%%%%%%%%%%%%%%%%%%%%%%%%%%%%%%%%%%%%%%%%%%%%%%%%
\subsection{X-ray luminosity during quiescence}
\label{pdp_lx}

One important observational property of GC CVs is
their X-ray luminosity. It is computed here as described in 
\citet[][see Section 3.3.4]{Belloni_2016a}. Most of the predicted 
CVs in our models have X-ray luminosities between $\sim 10^{29}$  
and $\sim 10^{30}$ erg s$^{-1}$. However, a few of them have 
$L_X$ below $10^{29}$ erg s$^{-1}$. These luminosities for our period-bouncers
seem to agree with the values found for GW Lib and WZ Sge, namely 
$0.05^{+0.10}_{-0.02}$ $\times 10^{30}$ and 
$0.7^{+0.3}_{-0.1}$ $\times 10^{30}$ erg s$^{-1}$ [2-10 keV], respectively 
\citep{Byckling_2010}. These systems are likely period-bouncers or CVs 
close to the period minimum, which makes them ideal for this comparison.

Even though our results seem to agree with the observed X-ray fluxes for CVs
close to the period minimum, recent investigations have found a correlation 
between the duty cycle and the X-ray luminosity for X-ray bright DNe in 
the solar neighborhood \citep{Britt_2015}. Our simulated CVs show a different
behaviour than this, most likely because the authors primarily have 
bright X-ray DNe in their sample ($L_X > 3 \times 10^{30}$ erg s$^{-1}$), whereas 
our simulated CVs are mostly period-bouncers or CVs close to the period minimum, 
with smaller X-ray luminosities. This makes a more rigorous comparison
difficult.

%%%%%%%%%%%%%%%%%%%%%%%%%%%%%%%%%%%%%%%%%%%%%%%%%%%%%%%%%%%%%%%%%%
% NEW SECTION
%%%%%%%%%%%%%%%%%%%%%%%%%%%%%%%%%%%%%%%%%%%%%%%%%%%%%%%%%%%%%%%%%%
\subsection{Magnitude during quiescence}
\label{pdp_mag}

Another important observational property of GC CVs is their magnitudes.  
Once a potential candidate is identified via its X-ray luminosity,
a confirmation via other techniques (such as identifying the optical
counterparts) are required in order to improve the
confidence level that the source is in fact a CV 
candidate\footnote{As in the Galactic field, 
only spectroscopy can confirm that a CV candidate is indeed
a CV \citep{Knigge_2012MMSAI}. However, since GCs are crowded fields, 
spectroscopy might be a challenge. The use of a 
combination of different techniques (H$\alpha$ and FUV 
imaging, X-ray, colour, late and negative superhumps, etc.) will therefore be needed to 
confirm the GC CV candidates, especially for DNe.}.

%NL: I added more clarification to the first sentence.  Please make sure it is right.
%DB: I replaced "accretion flow" with "stream of matter from the donor", 
%and "surface of the WD" with "accretion disk". Please check if it is okay.
In general, for period-bouncers, the optical flux is dominated
by the WD, with little (if any) contribution from the hot spot corresponding to
where the stream of matter from the donor strikes the accretion disk. 
This is especially true for the population of faint CVs found in the GC NGC 6397
\citep{Cohn_2010}. As claimed by these authors, the WD should be 
heated by the accretion process overall, since isolated massive WDs are 
faint enough to avoid detection (i.e., they have efficient cooling).

In our simulations, we estimate the CV magnitude as described
in \citet{Belloni_2016a}. No additional heating is added to the WDs.
For the cCAML and eCAML schemes, we do not have information
about the WD magnitude or temperature. Thus, we present our 
results only for the MOCCA formulation. Here, 
we have information about the WD magnitude and can compute the
CV magnitude by adding the flux contributions from the four main components: WD, donor,
hot spot and disk. 

Present-day CV magnitudes are strongly dependent on the time since CV 
formation, as well as on the formation mechanism. Basically, {\it the later the CV is 
formed (i.e., mass transfer starts), the brighter it is}. 
%DB: I changed the sentence below. Please check if it makes sense.
Since dynamically formed CVs are more massive, those newly formed CVs
are brighter than CVs newly formed from primordial binaries. 
Here, we define newly formed CVs as CVs formed after $\sim$ 10 Gyr of
cluster evolution, i.e. at most 2 Gyr ago.

%DB: I changed the sentence below. Please check if it makes sense.
Among those CVs that form at intermediate times (after $\sim$ 1 Gyr of cluster evolution 
but before  $\sim$ 10 Gyr), we see a transition in dynamically formed CVs; those formed a 
long time ago are currently fainter but those formed more recently are currently brighter 
than those CVs formed from primordial binaries. This transition occurs at $\sim$ 6 Gyr. 
The reason is twofold. On the one hand, this effect is partially due to 
the above-mentioned cooling efficiency of the WDs. But, on the other hand, 
the hot spot becomes the brightest CV component at this particular formation-time 
(and its luminosity is related to the donor mass, i.e. the higher the donor
mass, the higher the mass transfer rate, and in turn the higher the hot spot
luminosity). In other words, CVs formed at maximum $\sim$ 6 Gyr ago have their
optical fluxes dominated by their hot spots and those formed before this time
have their optical fluxes dominated by their WDs. This makes dynamical CVs
more luminous (currently) if formed after $\sim$ 6 Gyr of cluster evolution 
and CVs formed from primordial binaries more luminous (currently) if formed 
before $\sim$ 6 Gyr of cluster evolution.

For CVs that formed a long time ago (before $\sim$ 1 Gyr of cluster evolution), 
distinguishing between the different formation channels is more difficult.
%DB: I changed the sentence below. Please check if it makes sense.
This is because these CVs have more massive WDs irrespective of their formation 
channels, such that their WD cooling efficiencies and magnitudes are similar.

\begin{figure}
   \begin{center}
    \includegraphics[width=240 px]{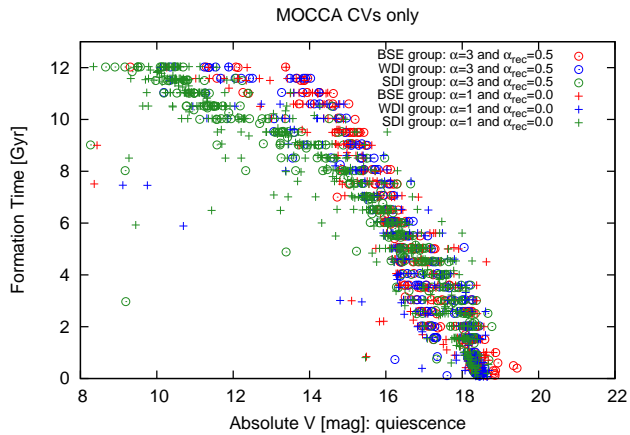} 
    \end{center}
  \caption{CV absolute V-band magnitude as a function of the CV formation time.  We plot all
CVs formed in MOCCA, and consider all 12 models.  These are separated according to our choices 
for the CEP parameters and our classifications for the influence of dynamics in CV formation. Note that, in general, dynamically 
formed CVs (SDI group) are brighter than CVs formed from primordial binaries (BSE and WDI groups).  This is the case 
provided they form close to the present-day (less than $\sim$ 6 Gyr ago).  At this point, namely $\sim$ 6 Gyr
ago, a transition occurs in which dynamically formed CVs start becoming fainter than CVs formed
from primordial binaries.  This is because, at this point, the CV flux starts to become dominated by the
WD flux. As pointed out in Section \ref{pdp_secondary_mass}, most CVs formed from primordial binaries have
He WDs, which are brighter than C/O WDs (due to the more efficient cooling in the latter). Finally, for CVs formed 
at $\sim$ 1 Gyr (or earlier), it is very difficult to distinguish between CVs formed from different formation channels.  
This is because, at these early times, basically all CVs have C/O WDs irrespective of the formation channel. 
%DB: I added "irrespective of the formation channel" above.
For more details see Section \ref{pdp_mag}.}
  \label{Fig05}
\end{figure}

The description above is illustrated in Fig. \ref{Fig05}, where only
CVs formed using the MOCCA scheme are shown. 
%All models with any indication of a dynamical influence during CV formation
%are shown, as well as our results for all CEP parameters. 
%DB: I replaced the commented above with the sentence below.
These are separated according to our choices for the CEP parameters and our 
classifications for the influence of dynamics in CV formation.
Note that the trends associated with each formation channel are independent of 
the CEP parameters.

We emphasize here that although we do not consider changes in the WD magnitude
arising from the accretion process, 
%DB: Please check the sentence below "we expect ... to be included". It might be my English issues, but seems confusing.
%NL:  I think the wording is fine, but another sentence could perhaps be added afterwards to include some justification for it (this 
% isn't obvious to me, at least).
%DB: I added something. Is it okay?
we expect the same overall results were this effect to be included. This is because
the above-described picture concentrates on the effects of dynamics in shaping present-day
GC CV magnitudes and detection limits.
This feature  or transitional path is similar in all models, as originally observed 
in the GC NGC 6397 \citep{Cohn_2010}.

%%%%%%%%%%%%%%%%%%%%%%%%%%%%%%%%%%%%%%%%%%%%%%%%%%%%%%%%%%%%%%%%%%
% NEW SECTION
%%%%%%%%%%%%%%%%%%%%%%%%%%%%%%%%%%%%%%%%%%%%%%%%%%%%%%%%%%%%%%%%%%
\subsection{Spatial distributions}
\label{pdp_spatial}

Finally, we provide a few words regarding the CV spatial distribution, as a 
function of the CV brightness. \citet{Cohn_2010} found
that there is no strong evidence in favour of the radial distribution 
of main-sequence-turn-off stars being different from the faint CVs 
($p$-value $\sim$ 0.04). However, for bright CVs, the evidence is 
sufficient ($p$-value $\sim$ 0.001) to make the claim that bright CVs are more 
centrally concentrated. 

In our models, we also typically find this trend.  However, it strongly 
depends on the following properties: (i) the 
source of energy in the host cluster, (ii) the host cluster evolution, (iii) the 
average mass in the host cluster core, (iv) the WD-MS binary and CV formation times, 
(v) the WD-MS binary and CV masses, and (vi) the formation channel
\citep[][see Section 5.7]{Belloni_2016a}. We emphasize
that the inclusion of the six additional models considered in this paper 
does not change 
the picture described in \citet{Belloni_2016a}, in particular 
CVs are most likely to be found between the core and half-mass radii.

%HERE!!!

%%%%%%%%%%%%%%%%%%%%%%%%%%%%%%%%%%%%%%%%%%%%%%%%%%%%%%%%%%%%%%%%%%
% NEW SECTION
%%%%%%%%%%%%%%%%%%%%%%%%%%%%%%%%%%%%%%%%%%%%%%%%%%%%%%%%%%%%%%%%%%
\section{DISCUSSION}
\label{discussion}

In the first paper of this series \citep{Belloni_2016a}, we presented 
six specific MOCCA models with a focus on the properties of
their present-day CV populations. In the second paper in this series \citep{Belloni_2016b}, 
we concentrated instead on the properties of the progenitor 
and formation-age populations, and the age-dependence of CV properties.  
In this paper, we extend these analyses further 
by considering twice the number of models, including two possible
combinations of the CEP parameters and two AML and CV evolution formulations.  
That is, our focus in this paper is on the influence of our adopted prescriptions 
for the binary evolution in determining our simulated CV properties, at different 
phases in the life of a CV (formation-age, present-day, etc.).
%NL: I added the above sentence for clarity.emphasis.  You could just put "binary evolution" in the title, 
%to make it shorter.  

In what follows, we discuss the principal
implications of our results with respect to the simulated GC CV properties.
We also discuss the dependences of our results on specific
assumptions and approaches.

%%%%%%%%%%%%%%%%%%%%%%%%%%%%%%%%%%%%%%%%%%%%%%%%%%%%%%%%%%%%%%%%%%
% NEW SECTION
%%%%%%%%%%%%%%%%%%%%%%%%%%%%%%%%%%%%%%%%%%%%%%%%%%%%%%%%%%%%%%%%%%
\subsection{Initial binary populations}
\label{discussion_CVs_ibp}

In this work, as in our previous papers, we simulate clusters that
follow two different IBPs (Section \ref{models}), namely the Kroupa 
and Standard IBPs. 
The main distributions associated with both IBPs 
are shown in Figures  \ref{A1} and \ref{A2} in the Appendix \ref{ap}.
%NL: I was thinking, it might be worth while adding in an Appendix, where you include a figure or two showing these different orbital
%parameter distributions.  Maybe a 2x2 (4 panel) figure showing the initial IBP periods, eccentricities, mass ratios, primary masses.  
%Something like that, so the reader has easy access to understanding how the distributions differ.
%DB: Great! :) The appendix was added. Please check if this is okay and if it corresponds to what you thought. 
%I added a reference to it in the last sentence above.
%NL: Looks Great!  Thanks for adding this!

The Standard IBP is associated with an uniform mass ratio distribution, an 
uniform distribution in the logarithm of the semi-major axis 
$\log(a)$ or a log-normal semi major axis distribution, and a thermal 
eccentricity distribution. In GC investigations, the Standard
IBP is usually adopted because it supplies hard binaries that
work as an energy source in the cluster over the long-term evolution. 
%NL: I don't follow this last sentence below.  Re-word?
%DB: I changed. Please take a look at it. Is it better? If not, we could simply remove, I think.
%It is just to make a connection between "Standard" and "Kroupa".
We note that in population synthesis codes for Galactic studies, 
the Standard IBP is also adopted, even though observational results show slightly different 
sorts of distributions. Such observational features are obtained directly
from the Kroupa IBP after some stimulated evolution (see below).
%NL: Looks good!  Only very minor re-wording from me.

It is usually accepted that the Galactic field population formed from the
dissolution of embedded star clusters \citep[e.g.][]{Lada_2003}  
after the expulsion of the residual gas during the star formation
process. In order to reconcile this picture with the observations of G and M-dwarf 
binaries in the Galactic field \citep{DM_1991,M_1992,FM_1992}, \citet{Kroupa_1995} developed
a theory in which star clusters experience some degree of dynamical processing, including disruption
of binaries, before dissolution. This dynamical processing should ultimately reproduce the 
observed properties of G and M-dwarfs in the Galactic field.  \citet{Kroupa_1995} also 
showed that a single dynamical operator corresponds to the mean embedded cluster 
from which the Galactic field G and M-dwarf binaries originate \citep{Marks_2012}.
The confirmation of this theory came later when data 
of all late-type binary systems near the Sun became available \citep{RG_1997}.

The Kroupa IBP has been tested against both numerical simulations and observations,
and has successfully explained the observational features of young clusters,
associations, and even binaries in old GCs
\citep[e.g.][and references therein]{Kroupa_2011,Marks_2012,Leigh_2015}.

In our investigation, our results show good agreement with the observed Galactic 
CV WD mass distribution for the Kroupa models.
However, the CV WD mass distribution is sensitive to AML prescriptions \citep{Schreiber_2016}
and more complete investigations should be carried out taking this into account.

%, even when the cCAML scheme is adopted. 
%MZ: What do you mean by ``even when the cCAML scheme is adopted''? What is strange or extreme in cCAML?
%DB: I removed "even when the cCAML scheme is adopted". The idea was to say that less He WD are produced
%if the Kroupa IBP is adopted, even in the cCAML scheme which is not supposed to remove CVs with He WD
%from the population.

Collectively, all of this suggests that the Kroupa IBP might be a better choice to
seed population synthesis codes (and GC simulations), compared to the Standard IBP.  However, a more thorough
investigation considering observational selection effects
and realistic Galactic star-forming processes (e.g. spatial distribution and 
stimulated evolution) needs to be conducted, 
including different constraints on, for example, Galactic WD-MS PCEBs
and CVs.

\begin{figure*}
   \begin{center}
    \includegraphics[width=0.49\linewidth]{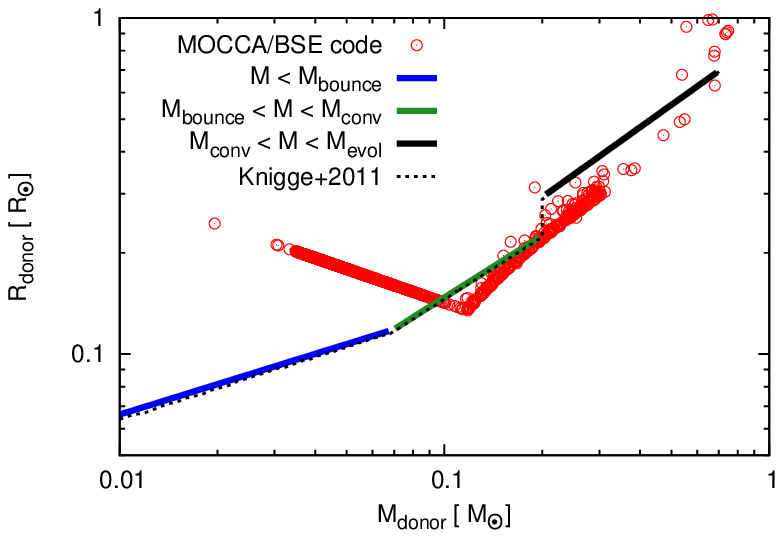} 
    \includegraphics[width=0.48\linewidth]{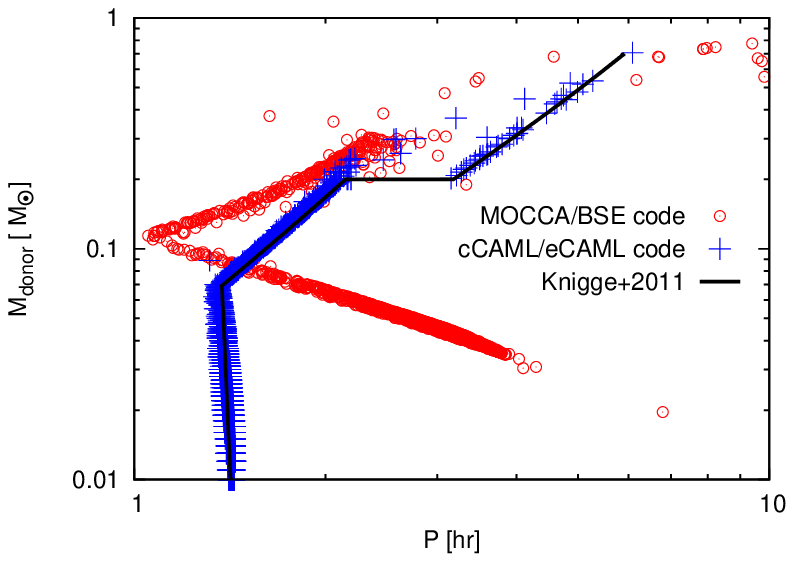} 
    \end{center}
  \caption{Donor mass-radius relation (left-hand panel) and period-donor mass relation (right-hand panel)
for both CV evolution schemes adopted in this work, namely MOCCA (or BSE code) and cCAML/eCAML (or the  state-of-the-art model,
i.e. the best-fit in \citet{Knigge_2011_OK} and further updates by \citet{Schreiber_2016}). We plot all present-day CVs formed
in all 12 models (for the MOCCA scheme) along with Eq. 16 and 17 in \citet{Knigge_2011_OK} in the
left-hand panels.  We plot all present-day CVs in all 12 models formed using all three binary evolution schemes
in the right-hand panels. Note that, due to outdated implementations in the BSE code, the CV evolution in these 
new simulations is 
drastically different than what we found in our previous set of simulations. For more details, see Section \ref{discussion_CVs_aml}.}
  \label{Fig06}
\end{figure*}

%%%%%%%%%%%%%%%%%%%%%%%%%%%%%%%%%%%%%%%%%%%%%%%%%%%%%%%%%%%%%%%%%%
% NEW SECTION
%%%%%%%%%%%%%%%%%%%%%%%%%%%%%%%%%%%%%%%%%%%%%%%%%%%%%%%%%%%%%%%%%%
\subsection{Influence of dynamics for different sets of initial cluster conditions}
\label{discussion_CVs_dynamics}

We saw in Section \ref{number_pdp_cvs_dynamics} that the dominant formation
channel in a cluster depends on its initial conditions. In general,
as expected, initially sparse clusters mainly form CVs similarly 
to the Galactic field, i.e. without any help
from dynamics.  On the other hand, when clusters are initially dense 
($\rho_c \gtrsim 10^4$ M$_\odot$ pc$^{-3}$), there is always a balance 
between dynamical creation and destruction of CV progenitors. 
This balance is strongly related to the cluster soft-hard 
boundary and the adopted IBP, and clearly outlining this 
dynamical picture might be challenging. Two important correlations, however, 
are: the importance of the CEP decreases as the cluster density increases, 
which was also found by \citet{Ivanova_2006}. The second correlation is found between 
the efficiency of the destruction of 
primordial CV progenitors and the initial cluster density; the higher
the initial density, the more primordial CV progenitors are destroyed.

\citet{Belloni_2016b} concluded that the CV formation rate could be 
similar in many clusters. As we saw in Section \ref{fap_formation_rate},
we rule out this initial impression, which was an artifact of 
small-number statistics. From the new models, we found that the CV formation
rate is not the same for all clusters, although CVs are continuously added to the population.  
After $\sim 1-2$ Gyr, the rate could still be nearly constant (with only the 
``constant of proportionality'' differing in different clusters).
%MZ: Again, is not better to use ``slope'' instead of ``linear coefficient''?
%DB: I replaced linear coefficient with "constant of proportionality".

Unfortunately, due to the limited sample of models analyzed here, we cannot reliably identify 
correlations between the initial or present-day cluster properties and the simulated CV properties, 
including the formation scenario, fraction of bright CVs, etc. 
This can be extended to include possible correlations between the CV formation rate and 
the initial cluster density or any present-day cluster properties. In future work, we aim to 
identify more correlations, by analyzing a larger subset of simulations from the 
MOCCA-SURVEY database \citep{Askar_2016b} with difference choices for the binary
stellar evolution parameters.

Finally, for the Kroupa models, \citet{Belloni_2016b} found that 
the net effect of weak dynamical interactions on the WDI group is strong. This is because
CVs in the WDI group had similar properties as those CVs strongly 
influenced of dynamics (SDI group). In this work, we find that the net effect
is weak (as in the Standard models), since CVs in the WDI
group have similar properties to those CVs in the BSE group (i.e., formed without
any influence from dynamics). This initial impression came from the fact that 
no CVs were found in the BSE group in our previous work (i.e., with larger efficiencies for the CEP). 
This is not the case when more realistic values of the CEP parameters are adopted.

%%%%%%%%%%%%%%%%%%%%%%%%%%%%%%%%%%%%%%%%%%%%%%%%%%%%%%%%%%%%%%%%%%
% NEW SECTION
%%%%%%%%%%%%%%%%%%%%%%%%%%%%%%%%%%%%%%%%%%%%%%%%%%%%%%%%%%%%%%%%%%
\subsection{CV evolution scheme}
\label{discussion_CVs_aml}

Given that the CV evolution model implemented in the BSE code (and used
in the MOCCA code) is outdated, we evolved separately all close WD-MS binaries 
formed in our simulations using the code described in \citep{Zorotovic_2016},
which is much more up-to-date. In this section, we discuss possible future improvements
that can be made to the BSE code for more realistic simulations.

Compared to more recent implementations, the main differences in the 
BSE prescriptions for CV evolution are:

\begin{enumerate}
\item the donor mass-radius relation adopted in BSE differs
drastically from that derived from observations, especially
in the regime of low-mass donors (M-dwarf and brown dwarf);
\item the normalization factors adopted in BSE for AML due to magnetic 
braking and gravitational radiation are different to those derived 
from the best-fit sequence by \citet{Knigge_2011_OK} adopted for cCAML
and renormalized for eCAML in the code from \citet{Zorotovic_2016};
\item the Roche-lobe radius definition used in BSE is that derived by
\citet{Eggleton_1983}, which represents the donor as a point source in 
the potential. In the state-of-the-art model, the donor Roche-lobe depends on the 
polytropic index of the donor;
%DB: I replaced "donor properties" with "donor Roche-lobe".
\item consequential angular momentum loss (CAML) that is associated 
with mass transfer (e.g. due to nova eruptions) is not considered in BSE;
\item the predicted donor radius does not increase in BSE once mass transfer 
begins. However, according to the observations, CVs above the orbital period 
gap are $\sim$ 30\% larger with respect to the radius they would have without 
mass transfer \citep[for more details][see Section 5.2]{Knigge_2011_OK}.
The code from \citet{Zorotovic_2016} uses the mass radius relation from 
\citet{Knigge_2011_OK} for CVs, which takes into account an expansion factor 
that allows us to reproduce the observed orbital period gap (which can not 
be reproduced with the BSE code). 
\item the critical mass ratio above which dynamical mass transfer from a 
low-mass MS star occurs is fixed to 0.695 in BSE. A more accurate critical mass
ratio estimate stems from equating the adiabatic mass-radius exponent to 
the donor's Roche-lobe mass-radius exponent, which depends strongly
on the adopted CAML scheme \citep[see][for more details]{Schreiber_2016}.
\end{enumerate}

In Fig. \ref{Fig06} we show both the donor mass-radius relations
and the CV evolution in the donor mass-period plane. 
The red dots in the left-hand panel correspond to all present-day CVs
simulated with BSE, while the dashed line is the donor mass-radius 
relation given by Eqs. 16 and 17 in \citet{Knigge_2011_OK} that is 
used for the cCAML/eCAML formulations. 
In the right-hand panel, we also included the donor mass versus orbital period
relation for all present-day CVs simulated with the code
from \citet{Zorotovic_2016} (blue crosses).
Notice how different the results from both models are, especially in the 
domain of extremely-low-mass donors ($\lesssim 0.1$ M$_\odot$). In the BSE 
code, CVs evolve rapidly towards longer periods after reaching the minimum 
period, which is not consistent with the observations.

Ultimately, all of the above differences should be corrected in BSE,
%MZ: I removed the ``easily'', because I'm not the only one who try to do this
%already, and it was not easy (that's indeed why we developed our own code)
%DB. Okay. Now I see. :) I removed "simplicity" from below as well.
which would allow for more realistic predictions for the subsequent 
CV evolution.  Upgrading the BSE code would be a service 
to the stellar dynamics community, since the BSE code is used in many star cluster 
evolution codes.

%%%%%%%%%%%%%%%%%%%%%%%%%%%%%%%%%%%%%%%%%%%%%%%%%%%%%%%%%%%%%%%%%%
% NEW SECTION
%%%%%%%%%%%%%%%%%%%%%%%%%%%%%%%%%%%%%%%%%%%%%%%%%%%%%%%%%%%%%%%%%%
\subsection{Common-envelope phase parameters}
\label{discussion_CVs_cep}

In this work, we simulate clusters with two choices for the 
CEP parameters (Section \ref{models}), namely
$\alpha$ = 3.0 and $\alpha_{\rm rec}$ = 0.5
and $\alpha$ = 1.0 and $\alpha_{\rm rec}$ = 0.0. 

One important, albeit alarming, result found by \citet{Belloni_2016a} is that
models following the Kroupa IBP do not produce CVs formed through
isolated binary evolution, when a high CEP efficiency is adopted. In this
work, we show that for more realistic values of the CEP parameters, 
this apparent problem is resolved. This is, in turn, an indication
that assuming high values for the CEP parameters combined with a Kroupa IBP is 
an inadequate combination. This is consistent with studies performed to date using 
the Standard IBP \citep{Zorotovic_2010,Toonen_2013,Camacho_2014}.

An important effect that arises by decreasing the CEP parameters is 
that the number of CVs increases, even though the number of PCEBs
decreases. The decrease in the number of PCEBs is caused by the
enhancement of mergers during the CEP. On the other hand, as the 
reduction in orbital energy is stronger for lower CEP efficiencies, 
systems that survive the CEP with a lower efficiency emerge from this
phase at shorter periods, i.e, they need less time to start mass transfer
and become CVs. This is seen in both IBPs adopted here. 

Finally, we note an increase in the number
of potential millisecond pulsars with He WD companions. Interestingly,
recent searches for near-ultraviolet counterparts to millisecond pulsars 
in the GC 47 Tuc have identified five such systems to date 
\citep{Rivera_2015}.

%%%%%%%%%%%%%%%%%%%%%%%%%%%%%%%%%%%%%%%%%%%%%%%%%%%%%%%%%%%%%%%%%%
% NEW SECTION
%%%%%%%%%%%%%%%%%%%%%%%%%%%%%%%%%%%%%%%%%%%%%%%%%%%%%%%%%%%%%%%%%%
\subsection{Consequential angular momentum loss treatment}
\label{discussion_CVs_caml}

In order to quantify the effects of CAML, which is not included in the BSE code, 
we consider in this investigation two formulations for CAML, namely the 
classical CAML (cCAML) from \citet{King_1995} and the empirical CAML 
(eCAML) from \citet{Schreiber_2016}, as described in Section \ref{CAML}.

The main effects on our results stem from the stability criterion 
for dynamical mass transfer, which strongly affects
the numbers and properties of our simulated CVs.

Let us first compare the criterion in the BSE code with the 
cCAML approach. As explained by \citet{Schreiber_2016}, the stability
criterion can be translated into a critical mass ratio $q_0$ above
which mass transfer is unstable (usually, dynamically or thermally). 
In the BSE code, the critical mass ratio for dynamically stable mass
transfer is fixed to $q_0 = 0.695$ for low-mass MS stars, 
%DB: I added that the above criterion is for low-mass MSs.
while the models that incorporate CAML derive a value for $q_0$ that depends on the donor mass. 
When cCAML is assumed, $q_0 \sim 0.91$ for donor masses below $\sim$ 0.4 M$_\odot$. 
Thereafter, it is a monotonically increasing function of the donor mass. 
However, although the cCAML formulation produces more CVs than the MOCCA models,
we see only minor differences upon comparing the WD mass distributions.

Next, we compare the cCAML and eCAML formulations. In the latter,
$q_0$ is strongly affected by the donor mass; it is a monotonically increasing 
function of the donor mass for all mass ranges \citep[][see their Fig. 2]{Schreiber_2016}.
Due to the way it was developed, eCAML removes all CVs whose WDs have He cores 
from the population due to dynamical mass transfer (which leads to merges).
The Kroupa models do not show strong variations upon adopting either the 
cCAML or eCAML schemes, except for the total absence of low-mass WDs and the 
smaller number of CVs with the eCAML scheme. 
This is because most CVs formed in the Kroupa models
have C/O WDs (with a small contribution from O/Ne/Mg WDs). 
These CVs are (presumably) weakly affected by 
enhanced AML as is provided by the eCAML scheme. 
For the Standard models, on the other hand, the WD mass distribution is strongly 
affected by the eCAML approach.

Based on our results, we emphasize that, in order to have better predictions
for GC CVs for comparisons to the observations, the more realistic 
cCAML or eCAML formulations should be used, along with more accurate CEP parameters.  
This should help to avoid neglecting the formation of particular CVs that might contribute
significantly to the global population, with the overall effects depending on the choice of IBP.

%%%%%%%%%%%%%%%%%%%%%%%%%%%%%%%%%%%%%%%%%%%%%%%%%%%%%%%%%%%%%%%%%%
% NEW SECTION
%%%%%%%%%%%%%%%%%%%%%%%%%%%%%%%%%%%%%%%%%%%%%%%%%%%%%%%%%%%%%%%%%%
\subsection{Empirical relations and assumptions}
\label{discussion_CVs_emp}

Throughout this work, we have relied on empirical
relations to compute important CV properties, such as the duration
of the outburst and its recurrence time. The most 
important such assumption is the donor mass-radius relation used in the 
cCAML and eCAML formulations. This is precisely what is shown in Fig. \ref{Fig06}.  
This relation was derived from observations of CV donors whose masses are
$\gtrsim$ 0.05 M$_\odot$. We saw in Section \ref{pdp_secondary_mass} that 
the majority of CVs formed using the cCAML and eCAML schemes have donor
masses below this value, independent of the formation channel. This
is, in principle, expected since GC CVs are much older than CVs in the 
solar neighborhood, and they have had enough time to evolve far beyond the
period minimum, at which point $M_2 (P_{\rm min}) \sim 0.07$ M$_\odot$. 
On the other hand, if the donor mass-radius relation for extremely-low-mass
BDs is somehow different, many CVs in our sample might have extremely low
mass transfer rates, which would make their detection next to impossible 
with current instrumentation. 
%This would drastically change our predictions,
%especially the relative numbers of faint and bright CVs.

The effects on the overall CV population due to extrapolations of the recurrence time \citep{Patterson_2011} 
and outburst duration \citep{Smak_1999} empirical relations should also be explored.  
Another way that might potentially change CV population properties is
the extrapolation of the empirical relations for recurrence times 
and durations of outbursts, in to the range of extremely-low-mass donors.
These expressions were derived from well-observed CVs in the Galaxy and could
be different for CVs far away from the period minimum. These relations are important
because they are used here to estimate CV duty cycles, which might be 
different if the empirical relations are different.

Most period-bouncers or period-bouncer progenitors undergo 
superoutbursts, with just a few (if any) normal outbursts occurring 
during supercycles.  Consequently, we might be missing important information 
that could lead to changes in our duty cycle estimates for period-bouncers.

%NL I don't think this sentence is need, really... Feel free to put it back and/or re-word it though.
%DB: Okay. I agree.
%Above all what was said, since it is not possible to retrieve precisely CV properties 
%from observations, especially for period-bouncers and period-bouncer progenitors, 
%we think our approach is still reasonable.

%%%%%%%%%%%%%%%%%%%%%%%%%%%%%%%%%%%%%%%%%%%%%%%%%%%%%%%%%%%%%%%%%%
% NEW SECTION
%%%%%%%%%%%%%%%%%%%%%%%%%%%%%%%%%%%%%%%%%%%%%%%%%%%%%%%%%%%%%%%%%%
\subsection{Comparisons to CV candidates in GCs}
\label{discussion_CVs_candidates}

Although rigorous comparisons between our models and real GC CVs are
difficult, due to strong degeneracies associated with the initial cluster 
conditions \citep{Askar_2016b}, general conclusions can still be drawn.  These are listed below.

We showed in Sections \ref{pdp_mag} and \ref{pdp_spatial} that bright CVs 
were mainly formed a few Gyr ago due to strong dynamical interactions
(mainly exchanges).  They tend to be more centrally concentrated, i.e. close to, although 
outside, the core. This is in agreement with what \citet{Cohn_2010} found
for their bright CVs. They suggested that CVs are born in (or close to)
the core and then migrate out of the core due to dynamical interactions, as they
age. In our simulations, we see that dynamical interactions are unlikely after
CV formation \citep{Belloni_2016b,Leigh_2016}.  Of those CVs that form 
in (or close to) the core and are kicked out to the outskirts, they quickly sink back in 
to the inner parts due to mass segregation.

This was also found by \citet{Hong_2016}, who investigated 81 simulated clusters with different 
initial masses, half-mass radii, Galactocentric distances, and primordial 
binary fractions.  The authors concluded that bright CVs (donor masses greater 
than 0.1 M$_\odot$) are more centrally concentrated than faint CVs. 
Additionally, the faint CVs have similar radial distributions relative 
to the main-sequence turn-off stars.

More generally, bright CVs are formed close to the core,
but are often kicked out of the core in the process, before finally
migrating back to the core due to mass segregation. Faint CVs can form (but not always) 
through strong dynamical interactions, close (or not) to the core. However,
these are highly evolved systems (much older than bright CVs) and, since 
they have masses that are similar to (or even smaller than) 
main-sequence turn-off stars, they should 
not be as centrally concentrated as bright CVs. What is important from 
an observational point of view is that most CVs (bright or faint) should be located
somewhere between the core and half-mass radii.

Our results suggest that detection through outbursts 
is possible for only a rather small fraction of the total CV population.  This is because 
their duty cycles are usually extremely small, especially using a more
realistic CAML treatment and CV evolution model. It follows that the apparent
lack of outbursts in GC CVs could just be a selection effect,
in the sense that our knowledge is limited to a small population of 
very close CVs that are frequently observed in outburst, since these 
are the easiest systems to detect \citep[e.g.][]{Servillat_2011,Knigge_2012MMSAI,Belloni_2016a}.
%, NL I did not follow this last sentence/phrase.  Reword?
%DB: It is okay without it, I think.
%instead of being caused by the fact that GC CVs are predominantly magnetic.

Finally, as has already been pointed out by \citet{Knigge_2012MMSAI}, 
the natural path toward improving our understanding of 
GC CVs is a deep survey for DNe in GCs, which would guarantee the
detection of at least a few WZ Sge systems. This would allow for a much more thorough 
comparison between the predictions of theory and observations, a crucial step toward disentangling the
true nature of GC CVs.

%%%%%%%%%%%%%%%%%%%%%%%%%%%%%%%%%%%%%%%%%%%%%%%%%%%%%%%%%%%%%%%%%%
% NEW SECTION
%%%%%%%%%%%%%%%%%%%%%%%%%%%%%%%%%%%%%%%%%%%%%%%%%%%%%%%%%%%%%%%%%%
%\section{Conclusions and Perspectives}
%\label{conclusions}
%DB: This section was included by Monica.
\section{CONCLUSIONS AND PERSPECTIVES}
\label{conclusions}

In terms of the WD mass distribution, we found that models that follow the Kroupa IBP show 
better agreement with the observed Galactic CVs than the those 
that assume the Standard IBP.
%DB: I replaced "model" with "IBP" and "IBP" with "model". 

Contrary to what we concluded in \citet{Belloni_2016b}, 
the CV formation rate seems to depend on our assumptions, especially during the 
first 1-2 Gyr of cluster evolution. Also, CVs formed through weak interactions (WDI group) seem to 
have similar properties to the ones formed without any interaction (BSE group), contrary 
to those CVs formed by strong interactions (SDI group).

Upon assuming more realistic (lower) values for the common envelope efficiencies, 
we found that more CVs (and less PCEBs) are produced, especially in the models 
that follow the Kroupa IBP.  Including the empirical approach for CAML from 
\citet{Schreiber_2016}, CVs with low-mass (helium-core) WDs are not produced, 
which is consistent with the observations.

Despite all the uncertainties involved in simulating CVs in GCs, we can infer from our simulations 
that bright CVs in GCs are young and mainly formed due to exchanges. 
These tend to be concentrated towards the centre of the cluster, but outside the 
core, and our simulations show that strong dynamical interactions become rare 
after the CV is formed. 

Due to the small duty cycles, especially when more
%DB: I replaced "short duration of their" with "small"
realistic models for CV evolution are used, our simulations suggest that 
only a small fraction of the total CV population should be detectable 
through outbursts. This would explain the apparent lack of outbursts in 
GC CVs.

%DB: I included the following plans to our work with regards to GC CVs and
%the Kroupa IBP.
Future investigations will concentrate on the influence of the CEP parameters 
as well as the CAML formalisms on populations of WD-MS PCEBs, as well as CVs formed from
the Kroupa IBP. For that purpose, we intend to carry 
out a detailed population synthesis study incorporating a realistic Galactic star-formation 
model (temporal and spatial) along with observational selection effects intended to 
further constrain stellar evolution parameters by direct comparisons to the 
observed properties of Galactic WD-MS PCEBs and CVs. Additionally, we aim to
analyse more GC models within the MOCCA-SURVEY to look for additional correlations between
GC and CV properties.

\section*{Acknowledgements}

DB was supported by the CAPES foundation, Brazilian Ministry of Education
through the grant BEX 13514/13-0 and by the National Science Centre
through the grant UMO-2016/21/N/ST9/02938.
MZ acknowledges financial support from FONDECYT (3130559).
MRS thanks for support from FONDECYT (1141269).
MG and AA were supported by
the National Science Centre through the grant DEC-2012/07/B/ST9/04412. 
AA would also like to acknowledge support from the National Science Centre 
through the grant UMO-2015/17/N/ST9/02573 and partial support from Nicolaus
Copernicus Astronomical Centre's grant for young researchers.

%%%%%%%%%%%%%%%%%%%%%%%%%%%%%%%%%%%%%%%%%%%%%%%%
%%%%%%  APPENDIX %%%%%%%%%%%%%%%%%%%%%%
%%%%%%%%%%%%%%%%%%%%%%%%%%%%%%%%%%%%%%%%%%%%%%%%

\appendix

\section[]{Initial Binary Populations}
\label{ap}

\begin{figure*}
   \begin{center}
    \includegraphics[width=0.75\linewidth]{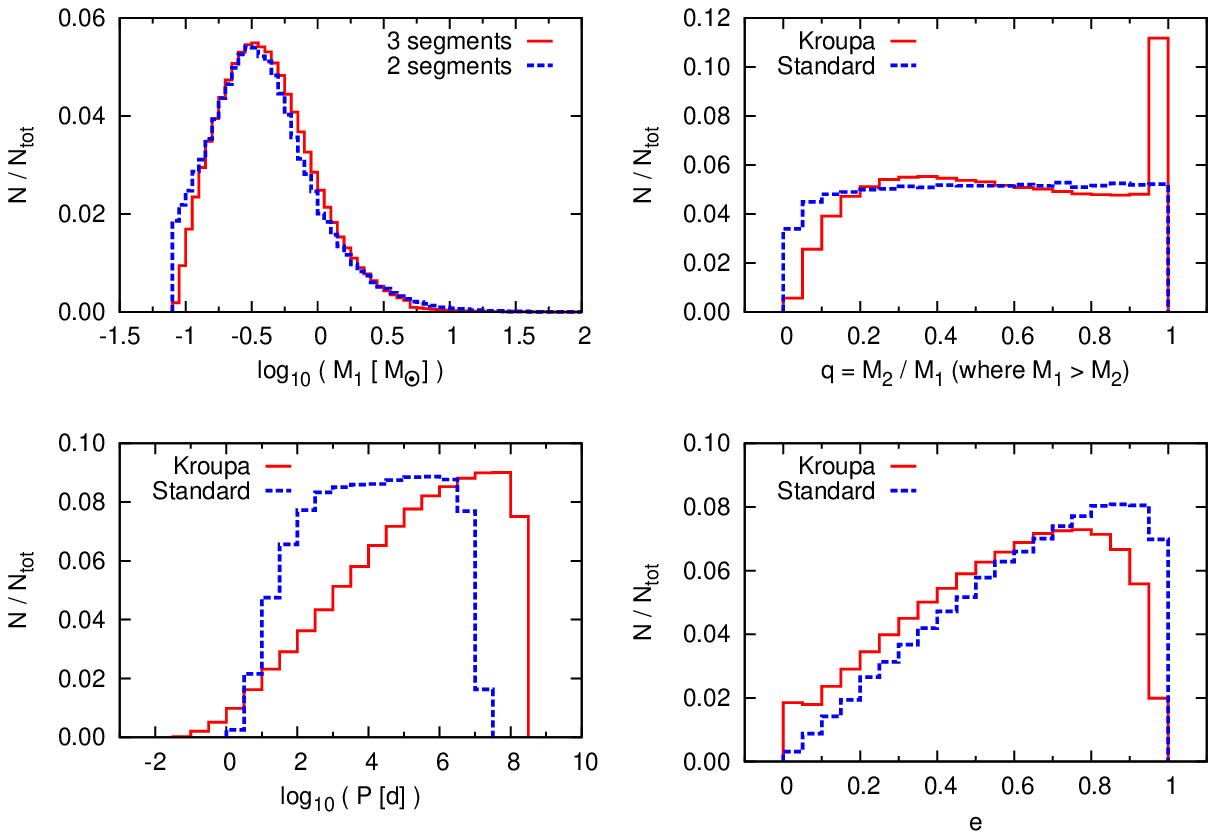} 
    \end{center}
  \caption{Primary mass (top left-hand panel), mass ratio (top right-hand panel), 
period (bottom left-hand panel), and eccentricity (bottom right-hand panel) distributions
for all initial binaries in models K2 (Kroupa IBP) and S2 (Standard IBP).}
  \label{A1}
\end{figure*}

In this investigation, we simulated models that follow two distinct initial binary populations (IBPs), 
namely: the Kroupa IBP and the Standard IBP (see Section \ref{models} for more details). In order to 
provide to the reader an easy way to recognize the differences between them, in Fig. \ref{A1}
we plot the main distributions associated with both IBPs for all initial binaries, i.e. primary 
mass (top left-hand panel), mass ratio (top right-hand panel), period (bottom left-hand 
panel), and eccentricity (bottom right-hand panel).

\begin{figure*}
   \begin{center}
    \includegraphics[width=0.75\linewidth]{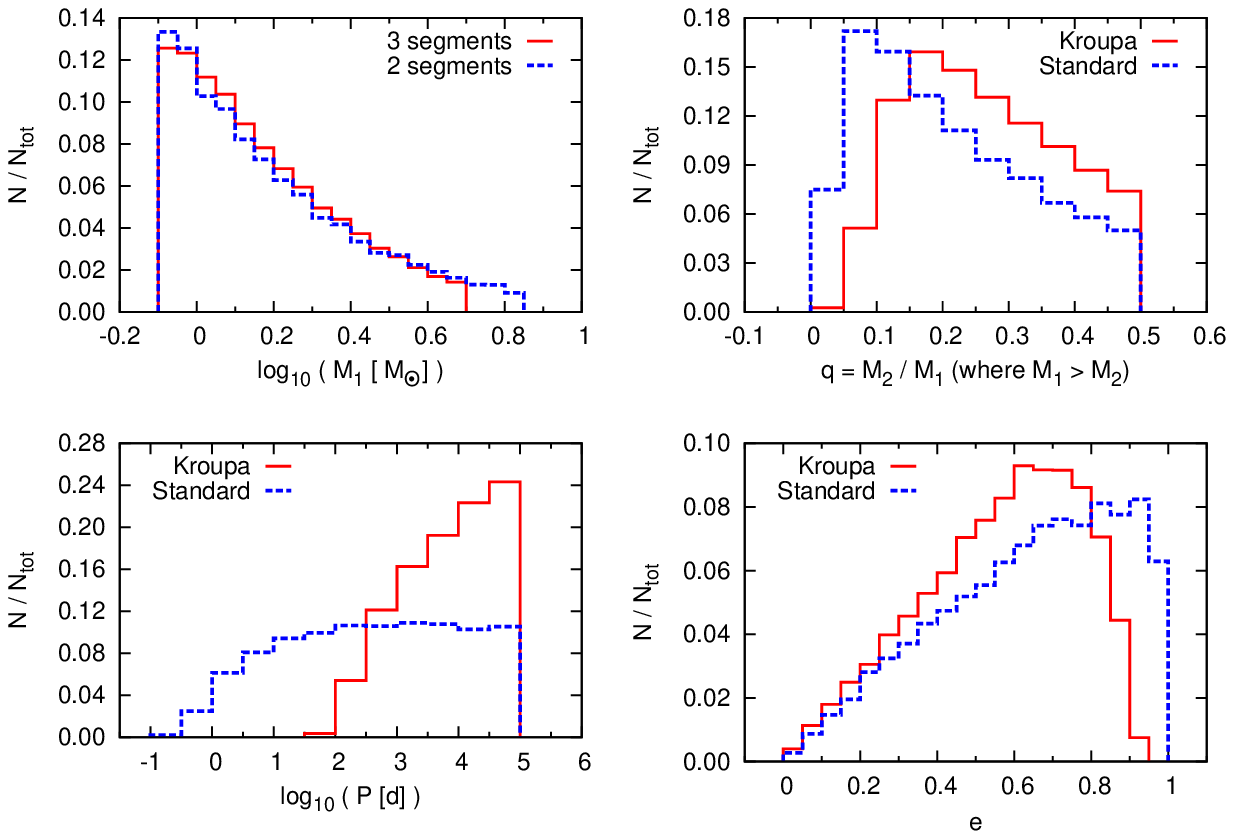} 
    \end{center}
  \caption{The same as in Fig. \ref{A1} but in the parameter space region defined by
$0.8 < M_1 < 7.0$ M$_\odot$, $\log_{10} (P/d) < 5.0$, and $q < 0.5$.  This range in the parameter space 
isolates the potential CV progenitors. Notice the differences between the two period 
distributions, which are significant.}
  \label{A2}
\end{figure*}

Additionally, since most CVs formed via pure binary stellar evolution come from initial binaries with 
primary masses between $\sim 0.8$ and $\sim 7.0$ M$_\odot$, periods shorter than $\sim 10^5$ d, 
and mass ratios smaller than 0.5, we plot in Fig. \ref{A2} the same distributions in Fig. \ref{A1}, 
but for the above-mentioned regions in the parameter space. This allows us to compare both the IBP 
for the entire population with the IBP of potential CV progenitors.

%%%%%%%%%%%%%%%%%%%%%%%%%%%%%%%%%%%%%%%%%%%%%%%%
%%%%%%  BIBLIOHRAPHYSTYLE %%%%%%%%%%%%%%%%%%%%%%
%%%%%%%%%%%%%%%%%%%%%%%%%%%%%%%%%%%%%%%%%%%%%%%%
\bibliographystyle{mnras}
\bibliography{references}

\begin{thebibliography}{}
\makeatletter
\relax
\def\mn@urlcharsother{\let\do\@makeother \do\$\do\&\do\#\do\^\do\_\do\%\do\~}
\def\mn@doi{\begingroup\mn@urlcharsother \@ifnextchar [ {\mn@doi@}
  {\mn@doi@[]}}
\def\mn@doi@[#1]#2{\def\@tempa{#1}\ifx\@tempa\@empty \href
  {http://dx.doi.org/#2} {doi:#2}\else \href {http://dx.doi.org/#2} {#1}\fi
  \endgroup}
\def\mn@eprint#1#2{\mn@eprint@#1:#2::\@nil}
\def\mn@eprint@arXiv#1{\href {http://arxiv.org/abs/#1} {{\tt arXiv:#1}}}
\def\mn@eprint@dblp#1{\href {http://dblp.uni-trier.de/rec/bibtex/#1.xml}
  {dblp:#1}}
\def\mn@eprint@#1:#2:#3:#4\@nil{\def\@tempa {#1}\def\@tempb {#2}\def\@tempc
  {#3}\ifx \@tempc \@empty \let \@tempc \@tempb \let \@tempb \@tempa \fi \ifx
  \@tempb \@empty \def\@tempb {arXiv}\fi \@ifundefined
  {mn@eprint@\@tempb}{\@tempb:\@tempc}{\expandafter \expandafter \csname
  mn@eprint@\@tempb\endcsname \expandafter{\@tempc}}}

\bibitem[\protect\citeauthoryear{{Askar}, {Szkudlarek}, {Gondek-Rosi{\'n}ska},
  {Giersz}  \& {Bulik}}{{Askar} et~al.}{2017}]{Askar_2016b}
{Askar} A.,  {Szkudlarek} M.,  {Gondek-Rosi{\'n}ska} D.,  {Giersz} M.,
  {Bulik} T.,  2017, \mn@doi [\mnras] {10.1093/mnrasl/slw177}, \href
  {http://adsabs.harvard.edu/abs/2017MNRAS.464L..36A} {464, L36}

\bibitem[\protect\citeauthoryear{{Belloni}, {Giersz}, {Askar}, {Leigh}  \&
  {Hypki}}{{Belloni} et~al.}{2016}]{Belloni_2016a}
{Belloni} D.,  {Giersz} M.,  {Askar} A.,  {Leigh} N.,   {Hypki} A.,  2016,
  \mn@doi [\mnras] {10.1093/mnras/stw1841}, \href
  {http://adsabs.harvard.edu/abs/2016MNRAS.462.2950B} {462, 2950}

\bibitem[\protect\citeauthoryear{{Belloni}, {Giersz}, {Rocha-Pinto}, {Leigh}
  \& {Askar}}{{Belloni} et~al.}{2017}]{Belloni_2016b}
{Belloni} D.,  {Giersz} M.,  {Rocha-Pinto} H.~J.,  {Leigh} N.~W.~C.,   {Askar}
  A.,  2017, \mn@doi [\mnras] {10.1093/mnras/stw2516}, \href
  {http://adsabs.harvard.edu/abs/2017MNRAS.464.4077B} {464, 4077}

\bibitem[\protect\citeauthoryear{{Britt} et~al.,}{{Britt}
  et~al.}{2015}]{Britt_2015}
{Britt} C.~T.,  et~al., 2015, \mn@doi [\mnras] {10.1093/mnras/stv256}, \href
  {http://adsabs.harvard.edu/abs/2015MNRAS.448.3455B} {448, 3455}

\bibitem[\protect\citeauthoryear{{Byckling}, {Mukai}, {Thorstensen}  \&
  {Osborne}}{{Byckling} et~al.}{2010}]{Byckling_2010}
{Byckling} K.,  {Mukai} K.,  {Thorstensen} J.~R.,   {Osborne} J.~P.,  2010,
  \mn@doi [\mnras] {10.1111/j.1365-2966.2010.17276.x}, \href
  {http://adsabs.harvard.edu/abs/2010MNRAS.408.2298B} {408, 2298}

\bibitem[\protect\citeauthoryear{{Camacho}, {Torres}, {Garc{\'{\i}}a-Berro},
  {Zorotovic}, {Schreiber}, {Rebassa-Mansergas}, {Nebot G{\'o}mez-Mor{\'a}n}
  \& {G{\"a}nsicke}}{{Camacho} et~al.}{2014}]{Camacho_2014}
{Camacho} J.,  {Torres} S.,  {Garc{\'{\i}}a-Berro} E.,  {Zorotovic} M.,
  {Schreiber} M.~R.,  {Rebassa-Mansergas} A.,  {Nebot G{\'o}mez-Mor{\'a}n} A.,
   {G{\"a}nsicke} B.~T.,  2014, \mn@doi [\aap] {10.1051/0004-6361/201323052},
  \href {http://adsabs.harvard.edu/abs/2014A%26A...566A..86C} {566, A86}

\bibitem[\protect\citeauthoryear{{Cannizzo} \& {Pudritz}}{{Cannizzo} \&
  {Pudritz}}{1988}]{Cannizzo_1988}
{Cannizzo} J.~K.,  {Pudritz} R.~E.,  1988, \mn@doi [\apj] {10.1086/166241},
  \href {http://adsabs.harvard.edu/abs/1988ApJ...327..840C} {327, 840}

\bibitem[\protect\citeauthoryear{{Claeys}, {Pols}, {Izzard}, {Vink}  \&
  {Verbunt}}{{Claeys} et~al.}{2014}]{Claeys_2014}
{Claeys} J.~S.~W.,  {Pols} O.~R.,  {Izzard} R.~G.,  {Vink} J.,   {Verbunt}
  F.~W.~M.,  2014, \mn@doi [\aap] {10.1051/0004-6361/201322714}, \href
  {http://adsabs.harvard.edu/abs/2014A%26A...563A..83C} {563, A83}

\bibitem[\protect\citeauthoryear{{Cohn} et~al.,}{{Cohn}
  et~al.}{2010}]{Cohn_2010}
{Cohn} H.~N.,  et~al., 2010, \mn@doi [\apj] {10.1088/0004-637X/722/1/20}, \href
  {http://adsabs.harvard.edu/abs/2010ApJ...722...20C} {722, 20}

\bibitem[\protect\citeauthoryear{{Dobrotka}, {Lasota}  \& {Menou}}{{Dobrotka}
  et~al.}{2006}]{Dobrotka_2006}
{Dobrotka} A.,  {Lasota} J.-P.,   {Menou} K.,  2006, \mn@doi [\apj]
  {10.1086/500042}, \href {http://adsabs.harvard.edu/abs/2006ApJ...640..288D}
  {640, 288}

\bibitem[\protect\citeauthoryear{{Duquennoy} \& {Mayor}}{{Duquennoy} \&
  {Mayor}}{1991}]{DM_1991}
{Duquennoy} A.,  {Mayor} M.,  1991, \aap, \href
  {http://adsabs.harvard.edu/abs/1991A%26A...248..485D} {248, 485}

\bibitem[\protect\citeauthoryear{{Eggleton}}{{Eggleton}}{1983}]{Eggleton_1983}
{Eggleton} P.~P.,  1983, \mn@doi [\apj] {10.1086/160960}, \href
  {http://adsabs.harvard.edu/abs/1983ApJ...268..368E} {268, 368}

\bibitem[\protect\citeauthoryear{{Fischer} \& {Marcy}}{{Fischer} \&
  {Marcy}}{1992}]{FM_1992}
{Fischer} D.~A.,  {Marcy} G.~W.,  1992, \mn@doi [\apj] {10.1086/171708}, \href
  {http://adsabs.harvard.edu/abs/1992ApJ...396..178F} {396, 178}

\bibitem[\protect\citeauthoryear{{Fregeau}, {Cheung}, {Portegies Zwart}  \&
  {Rasio}}{{Fregeau} et~al.}{2004}]{Fregeau_2004}
{Fregeau} J.~M.,  {Cheung} P.,  {Portegies Zwart} S.~F.,   {Rasio} F.~A.,
  2004, \mn@doi [\mnras] {10.1111/j.1365-2966.2004.07914.x}, \href
  {http://adsabs.harvard.edu/abs/2004MNRAS.352....1F} {352, 1}

\bibitem[\protect\citeauthoryear{{Fukushige} \& {Heggie}}{{Fukushige} \&
  {Heggie}}{2000}]{Fukushige_2000}
{Fukushige} T.,  {Heggie} D.~C.,  2000, \mn@doi [\mnras]
  {10.1046/j.1365-8711.2000.03811.x}, \href
  {http://adsabs.harvard.edu/abs/2000MNRAS.318..753F} {318, 753}

\bibitem[\protect\citeauthoryear{{Giersz}, {Heggie}, {Hurley}  \&
  {Hypki}}{{Giersz} et~al.}{2013}]{Giersz_2013}
{Giersz} M.,  {Heggie} D.~C.,  {Hurley} J.~R.,   {Hypki} A.,  2013, \mn@doi
  [\mnras] {10.1093/mnras/stt307}, \href
  {http://adsabs.harvard.edu/abs/2013MNRAS.431.2184G} {431, 2184}

\bibitem[\protect\citeauthoryear{{H{\'e}non}}{{H{\'e}non}}{1971}]{Henon_1971}
{H{\'e}non} M.~H.,  1971, Astrophysics and Space Science, 14, 151

\bibitem[\protect\citeauthoryear{{Hong}, {Vesperini}, {Belloni}  \&
  {Giersz}}{{Hong} et~al.}{2017}]{Hong_2016}
{Hong} J.,  {Vesperini} E.,  {Belloni} D.,   {Giersz} M.,  2017, \mn@doi
  [\mnras] {10.1093/mnras/stw2595}, \href
  {http://adsabs.harvard.edu/abs/2017MNRAS.464.2511H} {464, 2511}

\bibitem[\protect\citeauthoryear{{Hurley}, {Pols}  \& {Tout}}{{Hurley}
  et~al.}{2000}]{Hurley_2000}
{Hurley} J.~R.,  {Pols} O.~R.,   {Tout} C.~A.,  2000, \mn@doi [\mnras]
  {10.1046/j.1365-8711.2000.03426.x}, \href
  {http://adsabs.harvard.edu/abs/2000MNRAS.315..543H} {315, 543}

\bibitem[\protect\citeauthoryear{{Hurley}, {Tout}  \& {Pols}}{{Hurley}
  et~al.}{2002}]{Hurley_2002}
{Hurley} J.~R.,  {Tout} C.~A.,   {Pols} O.~R.,  2002, \mn@doi [\mnras]
  {10.1046/j.1365-8711.2002.05038.x}, \href
  {http://adsabs.harvard.edu/abs/2002MNRAS.329..897H} {329, 897}

\bibitem[\protect\citeauthoryear{{Ivanova}, {Heinke}, {Rasio}, {Taam},
  {Belczynski}  \& {Fregeau}}{{Ivanova} et~al.}{2006}]{Ivanova_2006}
{Ivanova} N.,  {Heinke} C.~O.,  {Rasio} F.~A.,  {Taam} R.~E.,  {Belczynski} K.,
    {Fregeau} J.,  2006, \mn@doi [\mnras] {10.1111/j.1365-2966.2006.10876.x},
  \href {http://adsabs.harvard.edu/abs/2006MNRAS.372.1043I} {372, 1043}

\bibitem[\protect\citeauthoryear{{Ivanova} et~al.,}{{Ivanova}
  et~al.}{2013}]{Ivanova_REVIEW}
{Ivanova} N.,  et~al., 2013, \mn@doi [\aapr] {10.1007/s00159-013-0059-2}, \href
  {http://adsabs.harvard.edu/abs/2013A%26ARv..21...59I} {21, 59}

\bibitem[\protect\citeauthoryear{{Ivanova}, {Justham}  \&
  {Podsiadlowski}}{{Ivanova} et~al.}{2015}]{Ivanova_2015}
{Ivanova} N.,  {Justham} S.,   {Podsiadlowski} P.,  2015, \mn@doi [\mnras]
  {10.1093/mnras/stu2582}, \href
  {http://adsabs.harvard.edu/abs/2015MNRAS.447.2181I} {447, 2181}

\bibitem[\protect\citeauthoryear{{King} \& {Kolb}}{{King} \&
  {Kolb}}{1995}]{King_1995}
{King} A.~R.,  {Kolb} U.,  1995, \mn@doi [\apj] {10.1086/175176}, \href
  {http://adsabs.harvard.edu/abs/1995ApJ...439..330K} {439, 330}

\bibitem[\protect\citeauthoryear{{Knigge}}{{Knigge}}{2012}]{Knigge_2012MMSAI}
{Knigge} C.,  2012, \memsai, \href
  {http://adsabs.harvard.edu/abs/2012MmSAI..83..549K} {83, 549}

\bibitem[\protect\citeauthoryear{{Knigge}, {Baraffe}  \& {Patterson}}{{Knigge}
  et~al.}{2011}]{Knigge_2011_OK}
{Knigge} C.,  {Baraffe} I.,   {Patterson} J.,  2011, \mn@doi [\apjs]
  {10.1088/0067-0049/194/2/28}, \href
  {http://adsabs.harvard.edu/abs/2011ApJS..194...28K} {194, 28}

\bibitem[\protect\citeauthoryear{{Kroupa}}{{Kroupa}}{1995}]{Kroupa_1995}
{Kroupa} P.,  1995, \mn@doi [\mnras] {10.1093/mnras/277.4.1507}, \href
  {http://adsabs.harvard.edu/abs/1995MNRAS.277.1507K} {277}

\bibitem[\protect\citeauthoryear{{Kroupa}}{{Kroupa}}{2008}]{Kroupa_INITIAL}
{Kroupa} P.,  2008, in {Aarseth} S.~J.,  {Tout} C.~A.,   {Mardling} R.~A.,
  eds,  Lecture Notes in Physics, Berlin Springer Verlag Vol. 760, The
  Cambridge N-Body Lectures. p.~181 (\mn@eprint {arXiv} {0803.1833}),
  \mn@doi{10.1007/978-1-4020-8431-7_8}

\bibitem[\protect\citeauthoryear{{Kroupa}}{{Kroupa}}{2011}]{Kroupa_2011}
{Kroupa} P.,  2011, in {Alves} J.,  {Elmegreen} B.~G.,  {Girart} J.~M.,
  {Trimble} V.,  eds,  IAU Symposium Vol. 270, Computational Star Formation. pp
  141--149 (\mn@eprint {arXiv} {1012.1596}), \mn@doi{10.1017/S1743921311000305}

\bibitem[\protect\citeauthoryear{{Kroupa}, {Gilmore}  \& {Tout}}{{Kroupa}
  et~al.}{1991}]{Kroupa_1991}
{Kroupa} P.,  {Gilmore} G.,   {Tout} C.~A.,  1991, \mnras, \href
  {http://adsabs.harvard.edu/abs/1991MNRAS.251..293K} {251, 293}

\bibitem[\protect\citeauthoryear{{Kroupa}, {Tout}  \& {Gilmore}}{{Kroupa}
  et~al.}{1993}]{Kroupa_1993}
{Kroupa} P.,  {Tout} C.~A.,   {Gilmore} G.,  1993, \mnras, \href
  {http://adsabs.harvard.edu/abs/1993MNRAS.262..545K} {262, 545}

\bibitem[\protect\citeauthoryear{{Lada} \& {Lada}}{{Lada} \&
  {Lada}}{2003}]{Lada_2003}
{Lada} C.~J.,  {Lada} E.~A.,  2003, \mn@doi [\araa]
  {10.1146/annurev.astro.41.011802.094844}, \href
  {http://adsabs.harvard.edu/abs/2003ARA%26A..41...57L} {41, 57}

\bibitem[\protect\citeauthoryear{{Lasota}}{{Lasota}}{2001}]{Lasota_2001}
{Lasota} J.-P.,  2001, \mn@doi [\nar] {10.1016/S1387-6473(01)00112-9}, \href
  {http://adsabs.harvard.edu/abs/2001NewAR..45..449L} {45, 449}

\bibitem[\protect\citeauthoryear{{Leigh}, {Giersz}, {Webb}, {Hypki}, {De
  Marchi}, {Kroupa}  \& {Sills}}{{Leigh} et~al.}{2013}]{Leigh_2013}
{Leigh} N.,  {Giersz} M.,  {Webb} J.~J.,  {Hypki} A.,  {De Marchi} G.,
  {Kroupa} P.,   {Sills} A.,  2013, \mn@doi [\mnras] {10.1093/mnras/stt1825},
  \href {http://adsabs.harvard.edu/abs/2013MNRAS.436.3399L} {436, 3399}

\bibitem[\protect\citeauthoryear{{Leigh}, {Giersz}, {Marks}, {Webb}, {Hypki},
  {Heinke}, {Kroupa}  \& {Sills}}{{Leigh} et~al.}{2015}]{Leigh_2015}
{Leigh} N.~W.~C.,  {Giersz} M.,  {Marks} M.,  {Webb} J.~J.,  {Hypki} A.,
  {Heinke} C.~O.,  {Kroupa} P.,   {Sills} A.,  2015, \mn@doi [\mnras]
  {10.1093/mnras/stu2110}, \href
  {http://adsabs.harvard.edu/abs/2015MNRAS.446..226L} {446, 226}

\bibitem[\protect\citeauthoryear{{Leigh}, {Geller}  \& {Toonen}}{{Leigh}
  et~al.}{2016}]{Leigh_2016}
{Leigh} N.~W.~C.,  {Geller} A.~M.,   {Toonen} S.,  2016, \mn@doi [\apj]
  {10.3847/0004-637X/818/1/21}, \href
  {http://adsabs.harvard.edu/abs/2016ApJ...818...21L} {818, 21}

\bibitem[\protect\citeauthoryear{{Marks} \& {Kroupa}}{{Marks} \&
  {Kroupa}}{2012}]{Marks_2012}
{Marks} M.,  {Kroupa} P.,  2012, \mn@doi [\aap] {10.1051/0004-6361/201118231},
  \href {http://adsabs.harvard.edu/abs/2012A%26A...543A...8M} {543, A8}

\bibitem[\protect\citeauthoryear{{Mazeh}, {Goldberg}, {Duquennoy}  \&
  {Mayor}}{{Mazeh} et~al.}{1992}]{M_1992}
{Mazeh} T.,  {Goldberg} D.,  {Duquennoy} A.,   {Mayor} M.,  1992, \mn@doi
  [\apj] {10.1086/172058}, \href
  {http://adsabs.harvard.edu/abs/1992ApJ...401..265M} {401, 265}

\bibitem[\protect\citeauthoryear{{Nelemans}, {Siess}, {Repetto}, {Toonen}  \&
  {Phinney}}{{Nelemans} et~al.}{2016}]{Nelemans_2016}
{Nelemans} G.,  {Siess} L.,  {Repetto} S.,  {Toonen} S.,   {Phinney} E.~S.,
  2016, \mn@doi [\apj] {10.3847/0004-637X/817/1/69}, \href
  {http://adsabs.harvard.edu/abs/2016ApJ...817...69N} {817, 69}

\bibitem[\protect\citeauthoryear{{Nomoto} \& {Kondo}}{{Nomoto} \&
  {Kondo}}{1991}]{Nomoto_1991}
{Nomoto} K.,  {Kondo} Y.,  1991, \mn@doi [\apjl] {10.1086/185922}, \href
  {http://adsabs.harvard.edu/abs/1991ApJ...367L..19N} {367, L19}

\bibitem[\protect\citeauthoryear{{Patterson}}{{Patterson}}{2011}]{Patterson_2011}
{Patterson} J.,  2011, \mn@doi [\mnras] {10.1111/j.1365-2966.2010.17881.x},
  \href {http://adsabs.harvard.edu/abs/2011MNRAS.411.2695P} {411, 2695}

\bibitem[\protect\citeauthoryear{{Rappaport}, {Verbunt}  \& {Joss}}{{Rappaport}
  et~al.}{1983}]{Rappaport_1983}
{Rappaport} S.,  {Verbunt} F.,   {Joss} P.~C.,  1983, \mn@doi [\apj]
  {10.1086/161569}, \href {http://adsabs.harvard.edu/abs/1983ApJ...275..713R}
  {275, 713}

\bibitem[\protect\citeauthoryear{{Reid} \& {Gizis}}{{Reid} \&
  {Gizis}}{1997}]{RG_1997}
{Reid} I.~N.,  {Gizis} J.~E.,  1997, \mn@doi [\aj] {10.1086/118436}, \href
  {http://adsabs.harvard.edu/abs/1997AJ....113.2246R} {113, 2246}

\bibitem[\protect\citeauthoryear{{Rivera-Sandoval} et~al.,}{{Rivera-Sandoval}
  et~al.}{2015}]{Rivera_2015}
{Rivera-Sandoval} L.~E.,  et~al., 2015, \mn@doi [\mnras]
  {10.1093/mnras/stv1810}, \href
  {http://adsabs.harvard.edu/abs/2015MNRAS.453.2707R} {453, 2707}

\bibitem[\protect\citeauthoryear{{Schenker}, {Kolb}  \& {Ritter}}{{Schenker}
  et~al.}{1998}]{Schenker_1998}
{Schenker} K.,  {Kolb} U.,   {Ritter} H.,  1998, \mn@doi [\mnras]
  {10.1046/j.1365-8711.1998.01529.x}, \href
  {http://adsabs.harvard.edu/abs/1998MNRAS.297..633S} {297, 633}

\bibitem[\protect\citeauthoryear{{Schmidtobreick}, {Shara}, {Tappert}, {Bayo}
  \& {Ederoclite}}{{Schmidtobreick} et~al.}{2015}]{Schmidtobreick_2015}
{Schmidtobreick} L.,  {Shara} M.,  {Tappert} C.,  {Bayo} A.,   {Ederoclite} A.,
   2015, \mn@doi [\mnras] {10.1093/mnras/stv250}, \href
  {http://adsabs.harvard.edu/abs/2015MNRAS.449.2215S} {449, 2215}

\bibitem[\protect\citeauthoryear{{Schreiber} et~al.,}{{Schreiber}
  et~al.}{2010}]{Schreiber_2010}
{Schreiber} M.~R.,  et~al., 2010, \mn@doi [\aap] {10.1051/0004-6361/201013990},
  \href {http://adsabs.harvard.edu/abs/2010A%26A...513L...7S} {513, L7}

\bibitem[\protect\citeauthoryear{{Schreiber}, {Zorotovic}  \&
  {Wijnen}}{{Schreiber} et~al.}{2016}]{Schreiber_2016}
{Schreiber} M.~R.,  {Zorotovic} M.,   {Wijnen} T.~P.~G.,  2016, \mn@doi
  [\mnras] {10.1093/mnrasl/slv144}, \href
  {http://adsabs.harvard.edu/abs/2016MNRAS.455L..16S} {455, L16}

\bibitem[\protect\citeauthoryear{{Servillat}, {Webb}, {Lewis}, {Knigge}, {van
  den Berg}, {Dieball}  \& {Grindlay}}{{Servillat}
  et~al.}{2011}]{Servillat_2011}
{Servillat} M.,  {Webb} N.~A.,  {Lewis} F.,  {Knigge} C.,  {van den Berg} M.,
  {Dieball} A.,   {Grindlay} J.,  2011, \mn@doi [\apj]
  {10.1088/0004-637X/733/2/106}, \href
  {http://adsabs.harvard.edu/abs/2011ApJ...733..106S} {733, 106}

\bibitem[\protect\citeauthoryear{{Shara} \& {Hurley}}{{Shara} \&
  {Hurley}}{2006}]{Shara_2006}
{Shara} M.~M.,  {Hurley} J.~R.,  2006, \mn@doi [\apj] {10.1086/504679}, \href
  {http://adsabs.harvard.edu/abs/2006ApJ...646..464S} {646, 464}

\bibitem[\protect\citeauthoryear{{Shara}, {Livio}, {Moffat}  \& {Orio}}{{Shara}
  et~al.}{1986}]{Shara_1986}
{Shara} M.~M.,  {Livio} M.,  {Moffat} A.~F.~J.,   {Orio} M.,  1986, \mn@doi
  [\apj] {10.1086/164762}, \href
  {http://adsabs.harvard.edu/abs/1986ApJ...311..163S} {311, 163}

\bibitem[\protect\citeauthoryear{{Shara} et~al.,}{{Shara}
  et~al.}{2007}]{Shara_2007}
{Shara} M.~M.,  et~al., 2007, \mn@doi [\nat] {10.1038/nature05576}, \href
  {http://adsabs.harvard.edu/abs/2007Natur.446..159S} {446, 159}

\bibitem[\protect\citeauthoryear{{Smak}}{{Smak}}{1999}]{Smak_1999}
{Smak} J.,  1999, \actaa, \href
  {http://adsabs.harvard.edu/abs/1999AcA....49..391S} {49, 391}

\bibitem[\protect\citeauthoryear{{Smak}}{{Smak}}{2001}]{Smak_2001}
{Smak} J.,  2001, in {L{\'a}zaro} F.~C.,  {Ar{\'e}valo} M.~J.,  eds,  Lecture
  Notes in Physics, Berlin Springer Verlag Vol. 563, Binary Stars: Selected
  Topics on Observations and Physical Processes. pp 110--150

\bibitem[\protect\citeauthoryear{{Stod{\'o}{\l}kiewicz}}{{Stod{\'o}{\l}kiewicz}}{1986}]{Stodolkiewicz_1986}
{Stod{\'o}{\l}kiewicz} J.~S.,  1986, \actaa, \href
  {http://adsabs.harvard.edu/abs/1986AcA....36...19S} {36, 19}

\bibitem[\protect\citeauthoryear{{Toonen} \& {Nelemans}}{{Toonen} \&
  {Nelemans}}{2013}]{Toonen_2013}
{Toonen} S.,  {Nelemans} G.,  2013, \mn@doi [\aap]
  {10.1051/0004-6361/201321753}, \href
  {http://adsabs.harvard.edu/abs/2013A%26A...557A..87T} {557, A87}

\bibitem[\protect\citeauthoryear{{Wang} et~al.,}{{Wang}
  et~al.}{2016}]{Wang_2016}
{Wang} L.,  et~al., 2016, \mn@doi [\mnras] {10.1093/mnras/stw274}, \href
  {http://adsabs.harvard.edu/abs/2016MNRAS.458.1450W} {458, 1450}

\bibitem[\protect\citeauthoryear{{Warner}}{{Warner}}{1995}]{Warner_1995_OK}
{Warner} B.,  1995, Cambridge Astrophysics Series, \href
  {http://adsabs.harvard.edu/abs/1995CAS....28.....W} {28}

\bibitem[\protect\citeauthoryear{{Webb} \& {Servillat}}{{Webb} \&
  {Servillat}}{2013}]{Webb_2013}
{Webb} N.~A.,  {Servillat} M.,  2013, \mn@doi [\aap]
  {10.1051/0004-6361/201117229}, \href
  {http://adsabs.harvard.edu/abs/2013A%26A...551A..60W} {551, A60}

\bibitem[\protect\citeauthoryear{{Willems}, {Kolb}, {Sandquist}, {Taam}  \&
  {Dubus}}{{Willems} et~al.}{2005}]{Willems_2005}
{Willems} B.,  {Kolb} U.,  {Sandquist} E.~L.,  {Taam} R.~E.,   {Dubus} G.,
  2005, \mn@doi [\apj] {10.1086/498010}, \href
  {http://adsabs.harvard.edu/abs/2005ApJ...635.1263W} {635, 1263}

\bibitem[\protect\citeauthoryear{{Yaron}, {Prialnik}, {Shara}  \&
  {Kovetz}}{{Yaron} et~al.}{2005}]{Yaron_2005}
{Yaron} O.,  {Prialnik} D.,  {Shara} M.~M.,   {Kovetz} A.,  2005, \mn@doi
  [\apj] {10.1086/428435}, \href
  {http://adsabs.harvard.edu/abs/2005ApJ...623..398Y} {623, 398}

\bibitem[\protect\citeauthoryear{{Zorotovic} \& {Schreiber}}{{Zorotovic} \&
  {Schreiber}}{2017}]{Zorotovic_2017}
{Zorotovic} M.,  {Schreiber} M.~R.,  2017, \mn@doi [\mnras]
  {10.1093/mnrasl/slw236}, \href
  {http://adsabs.harvard.edu/abs/2017MNRAS.466L..63Z} {466, L63}

\bibitem[\protect\citeauthoryear{{Zorotovic}, {Schreiber}, {G{\"a}nsicke}  \&
  {Nebot G{\'o}mez-Mor{\'a}n}}{{Zorotovic} et~al.}{2010}]{Zorotovic_2010}
{Zorotovic} M.,  {Schreiber} M.~R.,  {G{\"a}nsicke} B.~T.,   {Nebot
  G{\'o}mez-Mor{\'a}n} A.,  2010, \mn@doi [\aap] {10.1051/0004-6361/200913658},
  \href {http://adsabs.harvard.edu/abs/2010A%26A...520A..86Z} {520, A86}

\bibitem[\protect\citeauthoryear{{Zorotovic}, {Schreiber}  \&
  {G{\"a}nsicke}}{{Zorotovic} et~al.}{2011}]{Zorotovic_2011}
{Zorotovic} M.,  {Schreiber} M.~R.,   {G{\"a}nsicke} B.~T.,  2011, \mn@doi
  [\aap] {10.1051/0004-6361/201116626}, \href
  {http://adsabs.harvard.edu/abs/2011A%26A...536A..42Z} {536, A42}

\bibitem[\protect\citeauthoryear{{Zorotovic}, {Schreiber}  \&
  {Parsons}}{{Zorotovic} et~al.}{2014a}]{Zorotovic_2014}
{Zorotovic} M.,  {Schreiber} M.~R.,   {Parsons} S.~G.,  2014a, \mn@doi [\aap]
  {10.1051/0004-6361/201424430}, \href
  {http://adsabs.harvard.edu/abs/2014A%26A...568L...9Z} {568, L9}

\bibitem[\protect\citeauthoryear{{Zorotovic}, {Schreiber},
  {Garc{\'{\i}}a-Berro}, {Camacho}, {Torres}, {Rebassa-Mansergas}  \&
  {G{\"a}nsicke}}{{Zorotovic} et~al.}{2014b}]{Zorotovic_2014b}
{Zorotovic} M.,  {Schreiber} M.~R.,  {Garc{\'{\i}}a-Berro} E.,  {Camacho} J.,
  {Torres} S.,  {Rebassa-Mansergas} A.,   {G{\"a}nsicke} B.~T.,  2014b, \mn@doi
  [\aap] {10.1051/0004-6361/201323039}, \href
  {http://adsabs.harvard.edu/abs/2014A%26A...568A..68Z} {568, A68}

\bibitem[\protect\citeauthoryear{{Zorotovic} et~al.,}{{Zorotovic}
  et~al.}{2016}]{Zorotovic_2016}
{Zorotovic} M.,  et~al., 2016, \mn@doi [\mnras] {10.1093/mnras/stw246}, \href
  {http://adsabs.harvard.edu/abs/2016MNRAS.457.3867Z} {457, 3867}

\makeatother
\end{thebibliography}
%%%%%%%%%%%%%%%%%%%%%%%%%%%%%%%%%%%%%%%%%%%%%%%%

\bsp

\label{lastpage}

\end{document}